\documentclass[pra,aps,onecolumn]{revtex4-1}

\usepackage{amssymb,latexsym}
\usepackage{amsbsy} % for \boldsymbol
\usepackage{amsmath}
\usepackage{hyperref}
\usepackage{graphicx}
\usepackage[all]{xy}
\usepackage{color}

\newcommand \op {\hat}

\newcommand \bra[1] {\langle{#1}|}
\newcommand \ket[1] {|{#1}\rangle}
\newcommand \trace[1] {\mathrm{tr}({#1})}

\newcommand \pair[2] {({#1}|{#2})}
\newcommand \coulpair[2] {({#1}|r_{12}^{-1}|{#2})}

\newcommand \Tpara {\op{T}^{\text{para}}}

\newcommand \jpvec {\mathbf{j}_{\mathrm{p}}}

\newcommand \jmvec {\mathbf{j}_{\mathrm{m}}}

\newcommand \Aspace {\mathcal{A}}
\newcommand \Adiv {\mathcal{A}_{\mathrm{div}}}
\newcommand \Bdiv {L^2_{\mathrm{div}}}
\newcommand \Jdiv {\mathcal{J}_{\mathrm{div}}}

\newcommand \Estd {E}
\newcommand \Ecvx {\bar{E}}

\newcommand \FVR {f_{\mathrm{VR}}}
\newcommand \FVRcvx {\bar{f}_{\mathrm{VR}}}

\newcommand \FGH {F_{\mathrm{GH}}}
\newcommand \FGHcvx {\bar{F}_{\mathrm{GH}}}

\newcommand \EMS {E_{\mathrm{M}}}
\newcommand \EMScvx {\bar{E}_{\mathrm{M}}}

\newcommand \eMS {e_{\mathrm{M}}}
\newcommand \eMScvx {\bar{e}_{\mathrm{M}}}
\newcommand \FMS {F_{\mathrm{M}}}
\newcommand \FMSb {F_{\mathrm{M;b}}}
\newcommand \FMSj {F_{\mathrm{M;j}}}
\newcommand \FMScvx {\bar{F}_{\mathrm{M}}}
\newcommand \fMS {f_{\mathrm{M}}}
\newcommand \fMScvx {\bar{f}_{\mathrm{M}}}

\newcommand \commentout[1]  {}

\begin{document}

\title{Density-functional theory for internal magnetic fields}

\author{Erik I. Tellgren}
\email{erik.tellgren@kjemi.uio.no}
\affiliation{Hylleraas Centre for Quantum Molecular Sciences, Department of Chemistry, University of Oslo, N-0315 Oslo, Norway}

\begin{abstract}
A density-functional theory is developed based on the Maxwell--Schr\"odinger equation with an internal magnetic field in addition to the external electromagnetic potentials. The basic variables of this theory are the electron density and the total magnetic field, which can equivalently be represented as a physical current density. Hence, the theory can be regarded as a physical current-density functional theory and an alternative to the paramagnetic current density-functional theory due to Vignale and Rasolt. The energy functional has strong enough convexity properties to allow a formulation that generalizes Lieb's convex analysis-formulation of standard density-functional theory. Several variational principles as well as a Hohenberg--Kohn-like mapping between potentials and ground-state densities follow from the underlying convex structure. Moreover, the energy functional can be regarded as the result of a standard approximation technique (Moreau--Yosida regularization) applied to the conventional Schr\"odinger ground state energy, which imposes limits on the maximum curvature of the energy (w.r.t.\ the magnetic field) and enables construction of a (Fr\'echet) differentiable universal density functional.
\end{abstract}

\maketitle

\section{Introduction}

Density functional theory (DFT) is one of the most widely applied electronic structure methods in quantum chemistry and solid-state physics. This theory enables the standard problem of determining a ground-state wave function to be recast as the problem of determining only the total electron density of the ground state. Both the mathematical foundations and the density-functional approximations used in practice have reached a mature stage. Many attempts have been made to extend this computational framework to novel areas of application. One type of extension deals with Hamiltonians containing a magnetic vector potential representing an external magnetic field, which is beyond the scope of standard DFT. For example, the proof of the Hohenberg--Kohn theorem relies on the assumption that any two Hamiltonians differ by at most a multiplicative scalar potential~\cite{HOHENBERG_PR136_864}. Likewise, the constrained-search formulation~\cite{LEVY_PNAS76_6062} and Lieb's convex analysis formulation~\cite{LIEB_IJQC24_243,ESCHRIG03} also require modification to allow for external magnetic fields. The most conservative extensions to date are the magnetic-field density-functional theory (BDFT) due to Grayce and Harris~\cite{GRAYCE_PRA50_3089} as well as the paramagnetic current density functional theory (CDFT) due to Vignale and Rasolt~\cite{VIGNALE_PRL59_2360}. Less clear-cut are the attempts to formulate current density functional theories based on the physical current density due to Diener~\cite{DIENER_JP3_9417} as well as Pan and Sahni~\cite{PAN_IJQC110_2833}, which both suffer from gaps or mistakes in the proposed proofs of central results~\cite{VIGNALE_IJQC113_1422,TELLGREN_PRA86_062506,PAN_IJQC114_233,LAESTADIUS_IJQC114_782,LAESTADIUS_IJQC114_1445,LAESTADIUS_PRA91_032508}. Another possible route to incorporate external magnetic fields is via a relativistic density-functional theory based on the Dirac equation, which is likely to involve the physical current density as a basic variable. However, standard ground-state DFT relies on the variational principle and boundedness from below of the spectrum of the Schr\"odinger Hamiltonian, which are not available for the Dirac Hamiltonian. At present, the available formulations of relativistic ground-state DFT remain incomplete and non-rigorous~\cite{RAJAGOPAL_PRB7_1912,ESCHRIG_SSC56_777,ESCHRIG_JCC20_23,ENGEL_REST2002}. In particular, relativistic DFT is presently too incomplete to allow a clear non-relativistic limit to be constructed within the formalism. Such a limit, if it exists, is likely to be a type of physical CDFT.

In the present work, it will be shown how consideration of the energy of the induced magnetic field or, alternatively, a type of current-current interactions, allows a novel density-functional framework. This is done using the Maxwell--Schr\"odinger model, rather than the conventional Schr\"odinger equation, as the point of departure. The resulting DFT framework---termed MDFT~\footnote{The abbreviation MDFT stands for Maxwell--Schr{\"o}dinger-DFT or Moreau--Yosida-DFT}---can be regarded as a CDFT formulated in terms of the physical current density. Within the new formalism, the Grayce--Harris BDFT functional reappears in a different role. It will also be shown that the Maxwell--Schr\"odinger energy and related MDFT functionals have rich convexity properties and satisfy several variational principles.

The article is organized as follows. Basic concepts from convex analysis are briefly reviewed in Sec.~\ref{secCVXANL}. Next, in Sec.~\ref{secPARACDFT}, Grayce and Harris' BDFT formulation, and Vignale and Rasolt's paramagnetic CDFT formulations, both based on the conventional Schr\"odinger model, are reviewed along with basic definitions. In Sec.~\ref{secMS}, the Maxwell--Schr\"odinger model is specified. A density-functional theory is then formulated and analyzed, with focus on convexity properties. The main results appear in Sec.~\ref{secMDFT_LIEB}, \ref{secMDFT_MY}, and \ref{secMDFT_DIFF}. In Sec.~\ref{secUNIFORM}, a restriction of the theory to uniform magnetic fields is considered. The reduction from infinite-dimensional to low-dimensional representation of the magnetic field enables visualization of important aspects of the theory. Finally, a few concluding remarks are made in Sec.~\ref{secCONCL}.

\section{Review of convex analysis}
\label{secCVXANL}

Before turning to the formulation of density-functional theories, some useful concepts from convex analysis will be briefly reviewed---for a thorough exposition see Ref.~\onlinecite{ROCKAFELLAR09}.

Convex functions are an important well-characterized class of functions. In particular, this is one of the few large classes of functions for which global optimization is tractable. In many practical cases, it is sufficient to consider functions defined on a finite-dimensional domain. However, many results hold also for infinite-dimensional Banach spaces, a fact that proved useful in Lieb's reformulation of standard DFT~\cite{LIEB_IJQC24_243}. For the novel aspects of the present work, an intermediate level of generality is needed---it is sufficient to consider functions defined on a Hilbert space $H$.

A set $D \subset H$ is said to be convex if $x_1,x_2 \in D$, and $0 \leq \lambda \leq 1$, entails that $\lambda x_1 + (1-\lambda)x_2 \in D$. Geometrically, a line connecting any two points in a convex set is itself completely contained in the set. A function $f : D \to \mathbb{R}$, defined on a convex domain $D \subset H$, is said to be {\it convex} if
\begin{equation}
  f(\lambda x_1 + (1-\lambda) x_2) \leq \lambda f(x_1) + (1-\lambda) f(x_2), \quad 0 \leq \lambda \leq 1.
\end{equation}
Geometrically, this means that linear interpolation always overestimates the function. A function $f$ is said to be {\it concave} if $-f$ is convex, i.e.\ if linear interpolation instead underestimates the function. By choosing $x_1$ and $x_2$ to be global minimizers, it is seen that the global minima of a convex function form a convex set. This set can be empty, contain a unique point as the global minimum ($x_1=x_2$), or contain an infinite family of global minima. There can be no additional local minima besides the global minimum.

A function is said to be {\it affine} when the above relation is an equality, i.e.,
\begin{equation}
  f(\lambda x_1 + (1-\lambda) x_2) = \lambda f(x_1) + (1-\lambda) f(x_2), \quad 0 \leq \lambda \leq 1.
\end{equation}
Affine functions are both convex and concave. An important type of functions that are convex by construction are those that arise from maximization over a function $g: H\times A \to \mathbb{R}$ that is affine in the first argument:
\begin{equation}
  \label{eqMAXAFFPARAM}
  h(x) = \sup_{z\in A} g(x,z).
\end{equation}
Setting $x = \lambda  x_1 + (1-\lambda) x_2$, convexity follows directly,
\begin{equation}
  h(x) = \sup_{z} g(x,z) = \inf_{z} \big( \lambda g(x_1,z) + (1-\lambda) g(x_2,z) \big)
      \leq \lambda \inf_{y} g(x_1,z) + (1-\lambda) \inf_{z} g(x_2,z)
     = \lambda h(x_1) + (1-\lambda) h(x_2).
\end{equation}
With trivial adaptations, one also has that minimization over a function that is affine in the first arguments yields a concave function
\begin{equation}
  \label{eqMINAFFPARAM}
  h'(x) = \inf_{z\in A} g(x,z).
\end{equation}

A function $f : D \to \mathbb{R}$ is said to be {\it paraconvex} or {\it weakly convex} if there exists a $c > 0$ such that $f(x) + c \|x\|^2$ is convex. Equivalently, a paraconvex function satisfies
\begin{equation}
   f(\lambda x_1 + (1-\lambda) x_2)  \leq \lambda f(x_1) + (1-\lambda) f(x_2) + c \lambda (1-\lambda) \|x_1 - x_2\|^2.
\end{equation}
The definition of {\it paraconcave} functions is analogous. More generally, a function $f$ is said to be {\it difference convex} if there exists convex functions $g$ and $h$ such that $f = g-h$. By writing $f = (-h) - (-g)$ it is seen that a difference convex function is also {\it difference concave}.

A function $f : D \to \mathbb{R}$ is said to be {\it lower semicontinuous (l.s.c.)} if, for every $x_0$ and $\epsilon > 0$ there exists a neighborhood $U = \{ x \in D : \|x-x_0\| < \delta \}$ such that $f(x) \leq f(x_0) + \epsilon$ for all $x\in U$. A function $f$ is {\it upper semicontinuous (u.s.c.)} if $-f$ is l.s.c.

For a generic function $f(x)$, the {\it Legendre--Fenchel transformation} or {\it convex conjugate} is defined as
\begin{equation}
   f^*(y) = \sup_x \big( (y|x) - f(x) \big),
\end{equation}
where $(y|x)$ is the inner product of the Hilbert space. Noting that the right-hand side is affine in $y$, it can be seen to be a special case of Eq.~\eqref{eqMAXAFFPARAM} above. Hence, $f^*(y)$ is convex by construction. It can also be shown to be l.s.c.\ by construction. The double transform $f^{**}$ is equal to the original $f$ if and only if $f$ is both convex and l.s.c.

In the present context, it is convenient to change the sign convention in the Legendre--Fenchel transformation. We therefore define a Legendre--Fenchel transformation that maps convex functions to concave functions,
\begin{equation}
  f^{\wedge}(y) = \inf_x \big( (y|x) + f(x) \big),
\end{equation}
as well as a transformation that maps concave functions to convex functions,
\begin{equation}
  g^{\vee}(x) = \sup_y \big( g(x) - (y|x) \big).
\end{equation}
For a convex, l.s.c. function $f$, one then has $f = (f^{\wedge})^{\vee}$. Likewise, for a concave, u.s.c. function $g$, one has $g = (g^{\vee})^{\wedge}$.

A differentiability concept that is particularly well-adapted to convex functions is that of {\it subdifferentiability}. Given a point $x_0$ in the domain of a convex function $f$, a {\it subdifferential at $x_0$} is any element $y_0$ that satisfies
\begin{equation}
    f(x) \geq f(x_0) + y_0 (x-x_0),
\end{equation}
for all $x$. At points where the function is differentiable in the ordinary sense, the subdifferential is unique and equal to the ordinary derivative. At non-differentiable points, there may be multiple subdifferentials, a unique subdifferential, or none at all. The set of all subdifferentials is called the subgradient and is denoted
\begin{equation}
    \underline{\partial} f(x_0) = \{y_0 \, | \, f(x) \geq f(x_0) + y_0 (x-x_0) \ \text{for all} \ x\}.
\end{equation}
For concave functions, superdifferentials and supergradients are defined in the analogous way. That is, $y_0$ is a supergradient of the concave function $f(x_0)$ if and only if $y_0$ is a subgradient of the convex function $-f(x_0)$. The supergradient is denoted $\overline{\partial} f(x_0)$.

In convex analysis there is an operation that is somewhat analogous to convolution in Fourier analysis. The {\it infimal convolution} of two functions $f(x)$ and $g(x)$ is defined as
\begin{equation}
   (f \Box g)(x) = (g \Box f)(x) = \inf_{y} \big( f(x) + g(y-x) \big).
\end{equation}
When both $f$ and $g$ are convex l.s.c. functions, the Legendre--Fenchel transform takes the particularly simple form $(f \Box g)^* = f^* + g^*$. For general, non-convex functions one has $(f \Box g)^* \neq f^* + g^*$.

The infimal convolution often has regularizing effect. In the special case when $g_s(x) = \tfrac{1}{2s} \|x\|^2$, the infimal convolution $f\Box g_s$ is called the {\it Moreau--Yosida (MY) regularization} of $f$. In what follows, the notation
\begin{equation}
    M_s f(x) = (f\Box g_s)(x) = \inf_{y} \Big( f(y) + \frac{1}{2s} \|x-y\|^2 \Big)
\end{equation}
will be used. When $f$ is convex, the MY regularization indeed improves the regularity. For example, while $f$ need only be subdifferentiable, the MY regularization is always differentiable. On the other hand, when $f$ is non-convex, the function $M_s f$ is not guaranteed to be more regular. For example, it is possible that a differentiable, non-convex $f$ results in a non-differentiable $M_s f$. A connection between MY regularization and Legendre--Fenchel transformation is found by expanding the square in the above expression, leading to
\begin{equation}
    M_s f(x) = g_s(x) - (f+g_s)^*(\tfrac{1}{s} x).
\end{equation}
It follows immediately that $M_s f$ is paraconcave, that $M_s f - g_s$ is concave, and that the MY regularization is invertible whenever $f+g_s$ is convex and l.s.c.

An alternative to MY regularization that is more adapted to the non-convex case is the {\it Lasry--Lions (LL) regularization}~\cite{LASRY_IJM55_257,ATTOUCH_AIHP11_289}, defined by
\begin{equation}
   \label{eqLLREGDEF}
   L_{st} f(x) = -(g_t \Box -(g_s \Box f))(x)
        = \sup_{z} \inf_y \Big( f(y) + \frac{1}{2s} \|z-y\|^2  - \frac{1}{2t} \|x-z\|^2 \Big)
\end{equation}
and
\begin{equation}
   L^{st} f(x) = (g_t \Box -(g_s \Box -f))(x)
        = \inf_{z} \sup_y \Big( f(y) - \frac{1}{2s} \|z-y\|^2 + \frac{1}{2t} \|x-z\|^2 \Big).
\end{equation}
These operations can be identified with double MY regularizations---for example, $L_{st} f = -M_t(-M_s f)$. The first version results in approximations of $f(x)$ from below, whereas the second results in approximations from above. The approximation is furthermore tighter than a MY regularization:
\begin{equation}
   M_s f(x) \leq L_{st} f(x) \leq f(x), \quad s \geq t.
\end{equation}
When $s > t$, LL regularization of a quadratically bounded function, $f(x) \geq -\mathrm{const} \cdot (\|x\|^2 + 1)$, results in a Fr\'echet differentiable function with H\"older continuous derivative~\cite{ATTOUCH_AIHP11_289}. 

The degenerate case $s=t$ is known as the {\it proximal hull}~\cite{ROCKAFELLAR09}. Expanding the squares and reparametrizing the optimization of $z$ using $z' = z/s$ yields
\begin{align}
  L_{ss} f(x) & = -g_s(x) + (f+g_s)^{**}(x) \leq f(x),
           \\
  L^{ss} f(x) & = g_s(x) - (-f+g_s)^{**}(x) \geq f(x),
\end{align}
with equality when $f+g_s$ is convex and l.s.c. and $f-g_s$ is concave and u.s.c., respectively.

\section{Density-functional theory in the Schr\"odinger model}
\label{secPARACDFT}

At the non-relativistic level, an electronic system subject to
external electromagnetic potentials is typically described by the
Schr\"odinger equation, also called the Pauli equation when electron spin and the spin-Zeeman effect is explicitly considered. In SI-based atomic units, the ground-state energy is then given by
\begin{equation}
 \begin{split}
  \Estd(v,\mathbf{A}) & = \inf_{\psi} \bra{\psi} \op{T}_{\mathbf{A}} + \op{v} + \op{W} \ket{\psi} = \inf_{\Gamma} \trace{ \Gamma (\op{T}_{\mathbf{A}} + \op{v} + \op{W}) },
 \end{split}
\end{equation}
where the last minimization is performed over all valid mixed states $\Gamma$ and the Hamiltonian has been decomposed into three parts,
\begin{align}
   \op{T}_{\mathbf{A}} & = \frac{1}{2} \sum_l ( \boldsymbol{\sigma}_l\cdot \boldsymbol{\pi}_{\mathbf{A};l} )^2 = \frac{1}{2} \sum_l \big( \boldsymbol{\sigma}_l\cdot(\mathbf{p}_l+\mathbf{A}(\mathbf{r}_l)) \big)^2, \\
   \op{v} & = \sum_l v(\mathbf{r}_l), \\
   \op{W} & = \sum_{k<l} \frac{1}{r_{kl}},
\end{align}
where $\mathbf{p}_l = -i\nabla_l$ is the canonical momentum operator and $\boldsymbol{\pi}_{\mathbf{A}} = \mathbf{p} + \mathbf{A}$ the physical momentum operator. It will also prove useful to introduce a kinetic energy operator that
excludes the diamagnetic term,
\begin{equation}
   \Tpara_{\mathbf{A}} = \frac{1}{2} \sum_l ( \boldsymbol{\sigma}_l\cdot \mathbf{p}_l)^2 + \frac{1}{2} \sum_l\big\{ \boldsymbol{\sigma}_l\cdot \mathbf{p}_l, \, \boldsymbol{\sigma}_l\cdot\mathbf{A}(\mathbf{r}_l) \big\} = \op{T}_{\mathbf{A}} - \frac{1}{2} \sum_l |\mathbf{A}(\mathbf{r}_l)|^2.
\end{equation}

Any mixed state $\Gamma = \sum_k p_k \ket{\psi_k} \bra{\psi_k}$, where the weights satisfy $p_k \geq 0$ and $\sum_k p_k = 1$, gives rise to a one-particle reduced density matrix (1-RDM),
\begin{equation}
   \gamma(\mathbf{x}_1,\mathbf{x}_1') = \sum_k p_k \int \psi_k(\mathbf{x}_1,\mathbf{x}_2,\ldots,\mathbf{x}_N) \, \psi_k(\mathbf{x}'_1,\mathbf{x}_2,\ldots,\mathbf{x}_N)^* d\mathbf{x}_2 \cdots d\mathbf{x}_N,
\end{equation}
where $\mathbf{x}_j = (\mathbf{r}_j,\omega_j)$ contains both spatial and spin coordinates of particle $j$ and summation over spin degrees of freedom is implied. The 1-PRDM is diagonalized by the natural spinors (eigenvectors) $\phi_k$, with occupation numbers (eigenvalues) $n_k$,
\begin{equation}
   \gamma(\mathbf{x},\mathbf{x}') = \sum_k n_k \, \phi_k(\mathbf{x}) \, \phi_k(\mathbf{x}')^*.
\end{equation}
Organizing the natural spinors as two-component column vectors $\phi_k(\mathbf{r}) = (\phi^{\uparrow}_k(\mathbf{r}), \phi^{\downarrow}_k(\mathbf{r}))^T$, enables convenient expressions for the densities of interest in this work. The density and spin density are given by
\begin{align}
   \rho(\mathbf{r}) & = \sum_k n_k \phi_k(\mathbf{r})^{\dagger} \phi_k(\mathbf{r}), 
               \\
   \mathbf{m}(\mathbf{r}) & = \sum_k n_k \phi_k(\mathbf{r})^{\dagger} \boldsymbol{\sigma} \phi_k(\mathbf{r}).
\end{align}
Furthermore, the paramagnetic current density and magnetization current density are given by
\begin{align}
   \jpvec(\mathbf{r}) & = \frac{1}{2} \sum_k n_k \phi_k(\mathbf{r})^{\dagger} \mathbf{p} \phi_k(\mathbf{r}) + \text{c.c.}, 
               \\
   \jmvec(\mathbf{r}) & = \frac{1}{2} \sum_k n_k \phi_k(\mathbf{r})^{\dagger} \boldsymbol{\sigma} (\boldsymbol{\sigma}\cdot\mathbf{p}) \phi_k(\mathbf{r}) + \text{c.c.},
\end{align}
where `c.c.' denotes the complex conjugate of the preceeding expression. Both these current densities are gauge dependent and they differ by the spin current density $\jmvec - \jpvec = \nabla\times\mathbf{m}$. These currents can be contrasted with the gauge invariant, physical current density
\begin{equation}
   \label{eqJphysDef}
   \mathbf{j}_{\mathbf{A}}(\mathbf{r})  = \frac{1}{2} \sum_k n_k \phi_k(\mathbf{r})^{\dagger} \boldsymbol{\sigma} (\boldsymbol{\sigma}\cdot\boldsymbol{\pi}_{\mathbf{A}}) \phi_k(\mathbf{r}) + \text{c.c.}
\end{equation}
In what follows, the index $\mathbf{A}$ will sometimes be dropped but it should be noted that this current density explicitly depends on the external vector potential.

Standard density functional theory only takes into account the external scalar potential, and gives rise to a universal functional $F(\rho)$ and {\it the Hohenberg--Kohn variational principle},
\begin{equation}
    E(v) = \inf_{\rho\in X} \big( (\rho|v) + F(\rho) \big),
\end{equation}
where the pairing integral is defined by
\begin{equation}
  (\rho|v) = \int \rho(\mathbf{r}) \, v(\mathbf{r}) \, d\mathbf{r}.
\end{equation}
Different functionals $F$, with different convexity properties and different domains of densities they take finite values on, can be used in this variation principle. A unique $F$ is obtained from {\it the Lieb variational principle},
\begin{equation}
    F(\rho) = \sup_{v\in X^*} \big(E(v) -  (\rho|v)\big).
\end{equation}
Lieb showed that $\rho \in X = L^1 \cap L^3$, with the corresponding restriction $v \in X^* = L^{\infty} + L^{3/2}$ on the potentials, is a natural function space and minimization domain~\cite{LIEB_IJQC24_243}. Important aspects of Lieb's formulation include, first, that densities and potential belong to dual function spaces, i.e.\ the potential space is the space of all linear functionals defined on the density space via the pairing $(\rho|v)$. Second, that the energy functional $E(v)$ (in the absence of a vector potential) is concave in $v$ and that there exists a universal functional $F(\rho)$ in standard DFT that is convex in $\rho$, so that they can be related via Legendre--Fenchel transformations~\cite{LIEB_IJQC24_243,ESCHRIG03}. Third, concave (convex) functionals have either a unique global maximum (minimum) or a convex set of global maxima (minima), and no additional local maxima (minima). Fourth, though the functionals are not differentiable~\cite{LAMMERT_IJQC107_1943}, there is an applicable, rigorous notion of superdifferential for concave functionals and subdifferentials for convex functions. A potential $v_0$ with ground state density $\rho_0$ is then a subdifferential of $F(\rho_0)$ and $\rho_0$ is a superdifferential of $E(v_0)$. The Hohenberg--Kohn theorem can be reinterpreted in this setting as a mapping between super- and subdifferentials, and the conventional assumption of non-degenerate ground states can be dropped. Fifth, the Lieb variation principle provides a path to systematically approximate the exact functional $F(\rho)$ by introducing approximations into the maximization problem~\cite{COLONNA_JCP110_2828,FRYDEL_JCP112_5292,SAVIN_JCP115_6827,WU_JCP118_2498,TEALE_JCP130_104111,TEALE_JCP132_164115,TEALE_JCP133_164112,STROMSHEIM_JCP135_194109}.

External magnetic fields can be incorporated into density-functional theory in different ways. In BDFT due to Grayce and Harris, a semi-universal functional of the pair $(\rho,\mathbf{B})$ is introduced~\cite{GRAYCE_PRA50_3089}. As a slight modification of this idea, the semi-universal functional
\begin{equation}
  \label{eqFGHDef}
  \FGH(\rho,\mathbf{A}) = \inf_{\Gamma \mapsto \rho} \trace{ \Gamma (\op{T}_{\mathbf{A}} + \op{W}) } = \frac{1}{2} \pair{\rho}{A^2} + \inf_{\Gamma \mapsto \rho} \trace{ \Gamma (\Tpara_{\mathbf{A}} + \op{W}) },
\end{equation}
can be introduced. One can then write the BDFT variational principle
\begin{equation}
 \begin{split}
  \Estd(v,\mathbf{A}) & = \inf_{\rho\in X} \big( \pair{\rho}{v} + \FGH(\rho,\mathbf{A}) \big).
 \end{split}
\end{equation}
A complementary approach was introduced in by Vignale and Rasolt, who formulated current-density functional theory (CDFT) and defined a universal current density-dependent functional~\cite{VIGNALE_PRL59_2360,VIGNALE_PRB37_10685}. When the spin Zeeman term is neglected, the ground state energy functional can be expressed in terms of a universal functional of $(\rho,\jpvec)$. Subsequent developments identified an alternative formalism based on the magnetization current $\jmvec$~\cite{CAPELLE_PRL78_1872}. Rewriting the minimization as a nested minimization, with the inner minimization given by the Vignale--Rasolt constrained search functional,
\begin{equation}
 \begin{split}
   \FVR(\rho,\jmvec) = \inf_{\Gamma \mapsto \rho,\jmvec} \trace{ \Gamma (\op{T}_{\mathbf{0}} +\op{W}) }
 \end{split}
\end{equation}
yields
\begin{equation}
 \begin{split}
   \Estd(v,\mathbf{A})  & = \inf_{\substack{\rho \in X, \\ \jmvec \in Y}} \Big( (\rho|v+\tfrac{1}{2} A^2) + (\jmvec|\mathbf{A}) + \FVR(\rho,\jmvec) \Big).
 \end{split}
\end{equation}
For the current density and magnetic vector potential, a possible choice of function spaces is $\jmvec \in Y = L^1$ and $\mathbf{A} \in Y^* = L^{\infty}$~\cite{LAESTADIUS_IJQC114_1445}.  These function spaces guarantee a finite pairing $\pair{\jmvec}{\mathbf{A}}$ and a finite diamagnetic term $\frac{1}{2} \pair{\rho}{A^2}$. A subtle point is that in order for the functional $\FVR(\rho,\jmvec)$ to be {\it universal}, i.e.\ independent of the external potentials, the minimization domain of states $\Gamma$ must be universal too. For bounded vector potentials, this issue gets a simple solution, since the magnetic Sobolev space,
\begin{equation} 
  H^1_{\mathbf{A}} = \{ \psi \in L^2 \, | \, (\mathbf{p}_k+\mathbf{A}(\mathbf{r}_k)) \psi \in L^2 \},
\end{equation}
is independent of $\mathbf{A} \in L^{\infty}$ and equal to the regular Sobolev space $H^1 = \{ \psi \in L^2 \, | \, \mathbf{p}_k \psi \in L^2 \}$~\cite{LAESTADIUS_IJQC114_1445}. Hence, finiteness of the physical and canonical kinetic energies are equivalent, and a universal minimization domain, defined by the mixed states formed from pure states in $H^1$, can be used.

\begin{figure}[tb]
 \caption{\label{figpCDFTrel} Graphical representation of partial Legendre--Fenchel transforms, represented by double arrows, between functionals in the convex formulation of paramagnetic CDFT. Single arrows represent the the change of variables and additive shift by the diamagnetic energy that relate the cycle of partial transforms to the standard energy functional and the Grayce--Harris functional. The designations `concave-gen.' and `convex-gen.' are used for functionals that are generic in their second argument, i.e., do not have any apparent convexity properties.}
 \includegraphics[width=0.8\textwidth]{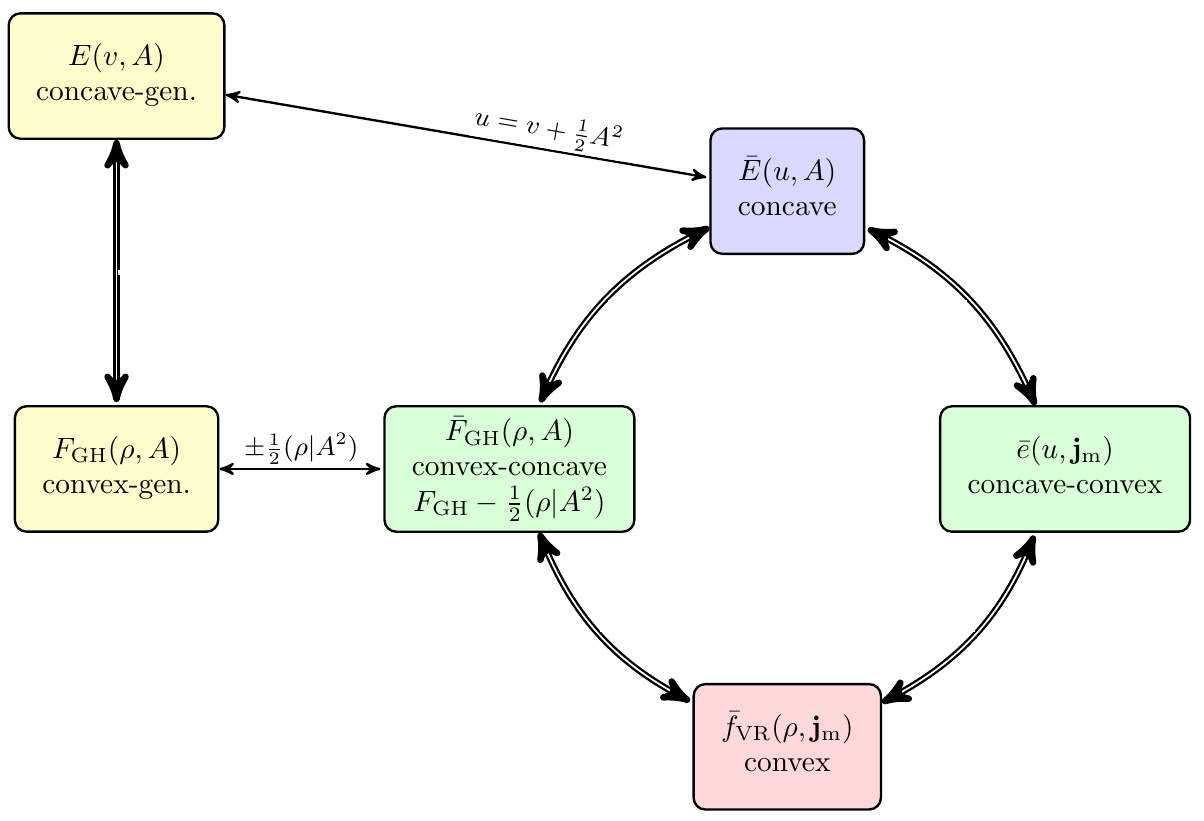}
\end{figure}

The concavity properties of $E(v,\mathbf{A})$ are of crucial importance for a convex analysis formulation along the lines of Lieb's formulation for standard DFT. However, while $E(v,\mathbf{A})$ is concave in $v$, it is not concave (nor convex) in $\mathbf{A}$. Relatedly, the semi-universal functional $\FGH(\rho,\mathbf{A})$ is convex in $\rho$ and non-concave in $\mathbf{A}$. For both functionals, the non-concavity is a manifestation of diamagnetism, which gives rise to \emph{local convexity} with respect to small perturbations of particular $(v,\mathbf{A})$ or $(\rho,\mathbf{A})$, for which the diamagnetic term is larger than the paramagnetic term. It would thus be desirable to be able to eliminate the diamagnetic energy. In fact, $\FGH(\rho,\mathbf{A})$ is {\it difference concave} in $\mathbf{A}$ and the difference, denoted
\begin{equation}
  \FGHcvx(\rho,\mathbf{A}) = \FGH(\rho,\mathbf{A}) - \frac{1}{2} \pair{\rho}{A^2} = \inf_{\jmvec \in Y} \big( \pair{\jmvec}{\mathbf{A}} + \FVR(\rho,\jmvec) \big),
\end{equation}
is manifestly concave in $\mathbf{A}$. This subtraction of the diamagnetic energy in the $(\rho,\mathbf{A})$-representation corresponds to the change of variables $u = v + \tfrac{1}{2} A^2$ in the $(v,\mathbf{A})$-representation. The resulting energy functional, $\Ecvx(u,\mathbf{A})$, is jointly concave and amendable to a convex analysis formulation that extends Lieb's formulation~\cite{TELLGREN_PRA86_062506}. Moreover, the universal functional $\FVR(\rho,\jmvec)$ is jointly convex in $(\rho,\jmvec)$. In the resulting theory, Legendre--Fenchel transforms connect the energy functional $\Ecvx(u,\mathbf{A})$ to the semi-universal BDFT functional and the universal CDFT functional~\cite{REIMANN_JCTC13_4089}:
\begin{equation}
 \begin{split}
  \Ecvx(u,\mathbf{A}) & = \inf_{\Gamma} \trace{ \Gamma (\Tpara_{\mathbf{A}} + \op{u} + \op{W}) } \\
    & = \inf_{\rho \in X} \big( \pair{\rho}{u} + \FGHcvx(\rho,\mathbf{A}) \big) = \inf_{\substack{\rho \in X, \\ \jmvec \in Y}} \big( \pair{\rho}{u} + \pair{\jmvec}{\mathbf{A}} + \FVR(\rho,\jmvec) \big).
 \end{split}
\end{equation}
Inversion of the last Legendre--Fenchel transformation yields
\begin{equation}
  \FVRcvx(\rho,\jmvec) = \sup_{\substack{u \in X^* \\ \mathbf{A} \in Y^*}} \big( \Ecvx(u,\mathbf{A}) - \pair{\rho}{u} - \pair{\jmvec}{\mathbf{A}} \big)
\end{equation}
At present, it is not known whether or not the constrained-search functional $\FVR$ is lower semicontinuous. If it is, then by the properties of Legendre--Fenchel transforms one has that $\FVRcvx$ and $\FVR$ are identical. If, however, $\FVR$ is not lower semicontinuous, it must be distinguished from $\FVRcvx$ which has this property by construction.  For completeness, one may similarly construct a concave-convex functional $\bar{e}(u,\jmvec)$,
\begin{equation}
  \bar{e}(u,\jmvec) = \inf_{\rho\in X} \big( \pair{\rho}{u} + \FVRcvx(\rho,\jmvec) \big) = \sup_{\mathbf{A} \in Y^*} \big( \Ecvx(u,\mathbf{A}) - \pair{\rho}{u} \big).
\end{equation}
The different transform pairs are summarized in Fig.~\ref{figpCDFTrel}.

Although the outlined theory has useful, well-defined convexity properties, the change of variables appears to be unnatural from the perspective that the non-relativistic CDFT is a limiting case of a relativistic density-functional theory based on the Dirac equation. At the relativistic level, no diamagnetic term $\tfrac{1}{2} A^2$ enters the Hamiltonian and it is presently unclear how the potential $u$ could arise in the non-relativistic limit. A related issue is how a universal functional of a gauge-dependent $\jmvec$ might arise in the non-relativistic limit of a theory that is formulated in terms of gauge invariant physical current densities. Relativistically, the decomposition of the physical current density into paramagnetic and diamagnetic parts is much less natural. These points merit further investigation but are beyond the scope of the present work. Instead, a very different convex formulation will be developed below.

\section{Density functional theory in a Maxwell--Schr\"odinger model}
\label{secMS}

The above Schr\"odinger model does not take into account that current densities associated with different electrons induce magnetic fields that affect the other electrons. Classically, a stationary current density $\mathbf{j}$ induces a magnetic field $\mathbf{b}$, as described by the magnetostatic Maxwell equation $\nabla \times \mathbf{b} = -\mu \mathbf{j}$. The energy of the induced field $\mathbf{b}$ is given by $\frac{1}{2 \mu} (\mathbf{b}|\mathbf{b}) = \frac{1}{2 \mu} \int |\mathbf{b}|^2 d\mathbf{r}$. In the present case, only magnetic fields with finite energy are of interest. A natural function space is thus the Hilbert space of square integrable functions, $\mathbf{b} \in L^2(\mathbb{R}^3)$. Every $L^2$ function can be represented as a Fourier transform, which will be denoted $\hat{\mathbf{b}}(\mathbf{k}) = \int e^{-i\mathbf{k}\cdot\mathbf{r}} \mathbf{b}(\mathbf{r}) \, d\mathbf{r}$. Adding the condition that a magnetic field is divergence free leads to the self-dual function space
\begin{equation}
 L^2_{\mathrm{div}}(\mathbb{R}^3) = \{\mathbf{b} \in L^2(\mathbb{R}^3) \, | \, \mathbf{k}\cdot\hat{\mathbf{b}} = 0\}.
\end{equation}
Two function spaces will be used for the magnetic vector potentials. The larger is simply
\begin{equation}
 \Aspace = \{\mathbf{a} \in L^1_{\mathrm{loc}} \, | \, k \, \hat{\mathbf{a}} \in L^2 \},
\end{equation}
where $k = |\mathbf{k}|$, and the smaller is the subset of divergence-free vector potentials,
\begin{equation}
 \Adiv = \{\mathbf{a} \in L^1_{\mathrm{loc}} \, | \, k \, \hat{\mathbf{a}} \in L^2, \ \mathbf{k}\cdot\hat{\mathbf{a}} = 0 \}.
\end{equation}
Here, the condition $\mathbf{a} \in L^1_{\mathrm{loc}}$ ensures that a Fourier transform $\hat{\mathbf{a}}$ exists at least in the distributional sense. For every magnetic field $\mathbf{b} \in \Bdiv$, one can \emph{define} a divergence-free current density using the above mentioned Maxwell equation. In terms of the Fourier transform, $\hat{\mathbf{j}} = -\mathbf{k} \times \hat{\mathbf{b}} / \mu$. The corresponding function space is
\begin{equation}
 \Jdiv = \{\mathbf{j} \in L^1_{\mathrm{loc}} \, | \, k^{-1} \, \hat{\mathbf{j}} \in L^2, \ \mathbf{k}\cdot\hat{\mathbf{j}} = 0 \}.
\end{equation}
A natural additional requirement on the current density space $\Jdiv$ would be that $\mathbf{j} \in L^1$ or $\hat{\mathbf{j}} \in L^{\infty}$, since this is true of the current densities computed from a wave function or mixed state. However, in order to be able to work with a self-dual space of magnetic fields, this route will not be taken. A restriction on $\Jdiv$ would result in a corresponding expansion of its dual space, and one would further need to distinguish between magnetic fields obtained as $\hat{\mathbf{b}} = -i \mu k^{-2} \mathbf{k} \times \hat{\mathbf{j}}$ and magnetic fields obtained as $\hat{\mathbf{b}}' = -i\mathbf{k}\times\hat{\mathbf{a}}$. Two such different fields would have a finite pairing $\pair{\mathbf{b}}{\mathbf{b}'}$ and the first type would have finite self-energy $\pair{\mathbf{b}}{\mathbf{b}}$, while the second self-energy $\pair{\mathbf{b}'}{\mathbf{b}'}$ would not be guaranteed to be finite.

For vector potentials $\mathbf{a} \in \Adiv$, the Maxwell equation
$-\nabla^2 \mathbf{a} = \nabla\times\mathbf{b} = -\mu \mathbf{j}$ can
be inverted to give
\begin{equation}
  \mathbf{a}(\mathbf{r}_1) = -\frac{\mu}{4\pi} \int \frac{\mathbf{j}(\mathbf{r}_2)}{|\mathbf{r}_1-\mathbf{r}_2|} d\mathbf{r}_2.
\end{equation}
Taking the curl of this expression yields Biot--Savart's law. Introducing the notation
\begin{equation}
   \coulpair{\mathbf{j}}{\mathbf{j}} = \int \frac{\mathbf{j}(\mathbf{r}_1)\cdot\mathbf{j}(\mathbf{r}_2)}{|\mathbf{r}_1-\mathbf{r}_2|} d\mathbf{r}_1 d\mathbf{r}_2
\end{equation}
and using partial integration establishes an inequivalence between magnetic field energy and current-current interaction,
\begin{equation}
    \frac{1}{2\mu} (\mathbf{b}|\mathbf{b}) = -\frac{1}{2} (\mathbf{j}|\mathbf{a}) = \frac{\mu}{8\pi} \coulpair{\mathbf{j}}{\mathbf{j}}.
\end{equation}

In what follows, the lower-case symbols $\mathbf{a},\mathbf{b},\mathbf{j}$ will be used to denote generic vector potentials, magnetic fields, and current densities. The symbol $\mathbf{j}$ will also be used to denote current densities computed from wave functions or mixed states. The external vector potential is denoted $\mathbf{A} \in \Aspace$, with associated magnetic field $\mathbf{B} =\nabla\times\mathbf{A}$ and current density $\mathbf{J} = -\nabla\times\mathbf{B} / \mu$. In addition, we introduce an internal field denoted by Greek letters: The internal vector potential is denoted $\boldsymbol{\alpha} \in \Aspace$, with associated magnetic field $\boldsymbol{\beta} = \nabla \times \boldsymbol{\alpha}$ and current density $\boldsymbol{\kappa} = -\nabla \times \boldsymbol{\beta} / \mu$. Inclusion of the energy in $\boldsymbol{\beta}$ and the mean-field interaction of $\boldsymbol{\alpha}$ with the electrons leads to the Maxwell--Schr\"odinger functional
\begin{equation}
 \begin{split}
   \label{eqEMSdef}
  \EMS(v,\mathbf{A}) & = \inf_{\substack{\psi \in H^1_{\boldsymbol{\alpha}+\mathbf{A}} \\ \boldsymbol{\alpha} \in \Aspace}} \Big( \frac{1}{2 \mu} (\boldsymbol{\beta}|\boldsymbol{\beta}) + \bra{\psi} \op{T}_{\boldsymbol{\alpha}+\mathbf{A}} + \op{v} + \op{W} \ket{\psi} \Big)
           \\
   & =  \inf_{\substack{\Gamma \\ \boldsymbol{\alpha} \in \Aspace}} \Big( \frac{1}{2 \mu} (\boldsymbol{\beta}|\boldsymbol{\beta}) + \trace{ \Gamma (\op{T}_{\boldsymbol{\alpha}+\mathbf{A}} + \op{v} + \op{W}) } \Big),
 \end{split}
\end{equation}
where $H^1_{\boldsymbol{\alpha}+\mathbf{A}}$ is a magnetic Sobolev space and the infimum over $\Gamma$ is taken over all corresponding mixed states. It will be seen below that this leads to a universal functional, despite the appearance of the external vector potential in $H^1_{\boldsymbol{\alpha}+\mathbf{A}}$. This functional can be regarded as a simplified mean-field picture of the radiation degrees of freedom considered by Ruggenthaler and co-workers~\cite{RUGGENTHALER_PRA84_042107,RUGGENTHALER_PRA90_012508,RUGGENTHALER_JP27_203202} in work combining non-relativistic quantum electrodynamics and time-dependent density functional theory. The stationarity conditions for the Maxwell--Schr\"odinger functional are a set of coupled equations: The energy eigenvalue equation for a Schr\"odinger Hamiltonian,
\begin{equation}
   \label{eqMSStatCond1}
  \Big( \frac{1}{2 \mu} (\boldsymbol{\beta}|\boldsymbol{\beta}) + \op{T}_{\boldsymbol{\alpha}+\mathbf{A}} + \op{v} + \op{W} \Big) \ket{\psi} = \EMS(v,\mathbf{A}) \ket{\psi_{\mathrm{M}}},
\end{equation}
and one of Maxwell's equations (the magnetostatic specialization of Amp\'ere's circuital law),
\begin{equation}
   \label{eqMSStatCond2}
  \frac{1}{\mu} \nabla\times\boldsymbol{\beta}(\mathbf{r}) + \mathbf{j}_{\boldsymbol{\alpha}+\mathbf{A}}(\mathbf{r}) = 0,
\end{equation}
where $\mathbf{j}_{\boldsymbol{\alpha}+\mathbf{A}}$ is the physical current density arising from the state $\psi_{\mathrm{M}}$ and the total vector potential. The self-consistent nature of the condition is more apparent when the physical current density is decomposed into diamagnetic and paramagnetic terms,
\begin{equation}
  \frac{1}{\mu} \nabla\times\boldsymbol{\beta}(\mathbf{r}) + \rho(\mathbf{r}) \, \boldsymbol{\alpha}(\mathbf{r}) + \rho(\mathbf{r}) \, \mathbf{A}(\mathbf{r})  + \jmvec(\mathbf{r}) = 0,
\end{equation}
where $\rho$ is the density resulting from $\psi_{\mathrm{M}}$.

Some further insight into the model can be gained by rewriting the ground-state energy in a form that involves magnetostatic current-current interactions. Using the identity $\op{T}_{\boldsymbol{\alpha}+\mathbf{A}} = \op{T}_{\mathbf{A}} + \frac{1}{2}\{\boldsymbol{\pi}_{\mathbf{A}}, \boldsymbol{\alpha}\} + \frac{1}{2} \alpha^2$ one obtains
\begin{equation}
   \EMS(v,\mathbf{A}) = \frac{(\boldsymbol{\beta}|\boldsymbol{\beta})}{2\mu} + \bra{\psi_{\mathrm{M}}} \op{T}_{\boldsymbol{\alpha}+\mathbf{A}}+\op{v}+\op{W} \ket{\psi_{\mathrm{M}}}  = \frac{\mu}{8\pi} \coulpair{\boldsymbol{\kappa}}{\boldsymbol{\kappa}} + (\mathbf{j}_{\mathbf{A}} + \tfrac{1}{2} \rho \boldsymbol{\alpha}|\boldsymbol{\alpha}) + \bra{\psi_{\mathrm{M}}} \op{T}_{\mathbf{A}}+\op{v}+\op{W} \ket{\psi_{\mathrm{M}}},
\end{equation}
where $\mathbf{j}_{\mathbf{A}} = \mathbf{j}_{\boldsymbol{\alpha}+\mathbf{A}} - \rho\boldsymbol{\alpha}$ is the current density computed using only the external vector potential. Assume now that the gauge of $\psi_{\mathrm{M}}$ is such that $\boldsymbol{\alpha}$ is divergence free. Then the self-consistency condition $\boldsymbol{\kappa} = \mathbf{j}_{\boldsymbol{\alpha}+\mathbf{A}}$ can be used to first rewrite the field energy and second to express $\boldsymbol{\alpha}$ using Biot--Savart's law. The result,
\begin{equation}
   \EMS(v,\mathbf{A}) = -\frac{\mu}{8\pi} \coulpair{\mathbf{j}_{\mathbf{A}}}{\mathbf{j}_{\boldsymbol{\alpha}+\mathbf{A}}} + \bra{\psi_{\mathrm{M}}} \op{T}_{\mathbf{A}}+\op{v}+\op{W} \ket{\psi_{\mathrm{M}}},
\end{equation}
shows that the Maxwell--Schr\"odinger energy implicitly includes a type of current-current interaction between $\mathbf{j}_{\boldsymbol{\alpha}+\mathbf{A}}$ and $\mathbf{j}_{\mathbf{A}}$.

Although the Maxwell--Schr\"odinger energy is different from the Schr\"odinger energy, the two quantities are related in several ways. For example, the infimum over $\boldsymbol{\alpha}$ can be overestimated by setting $\boldsymbol{\alpha} = 0$ or $\boldsymbol{\alpha} = -\mathbf{A}$. It immediately follows that
\begin{align}
  \EMS(v,\mathbf{A}) & \leq \Estd(v,\mathbf{A}),       \label{eqEMSineqE}
            \\
  \EMS(v,\mathbf{A}) & \leq \Estd(v,\mathbf{0}) + \frac{1}{2\mu} (\mathbf{B}|\mathbf{B}).          \label{eqEMSineqG}
\end{align}
These inequalities show that the Maxwell--Schr\"odinger ground state functional is a general a lower bound on the Schr\"odinger ground state functional and that its magnetic-field variation is bounded by the energy of the external magnetic field.
Overestimating the infimum in the definition of $\EMS(v,0)$ by taking $\boldsymbol{\alpha} = \mathbf{A}$ similarly leads to
\begin{equation}
  \label{eqPARAMAGBOUND}
   \EMS(v,0) \leq \frac{1}{2\mu} (\mathbf{B}|\mathbf{B}) + \inf_{\Gamma} \Big( \trace{ \Gamma (\op{T}_{\mathbf{A}} + \op{v} + \op{W}) } \Big) = \frac{1}{2\mu} (\mathbf{B}|\mathbf{B}) + E(v,\mathbf{A}),
\end{equation}
setting a bound on how strongly paramagnetic a system can be.

The infimum defining $\EMS(v,\mathbf{A})$ can also be overestimated by fixing $\psi$ to be the ground state of the Schr\"odinger Hamiltonian $T_{\mathbf{A}}+v+W$. Let $\mathbf{j}_{\mathrm{S}}$ be the corresponding current density evaluated using only the external vector potential. Finally, the infimum over internal vector potential can be overestimated by setting
\begin{equation}
   \boldsymbol{\alpha}(\mathbf{r}_1) = -\frac{\lambda}{4\pi} \int \frac{\mathbf{j}_{\mathrm{S}}(\mathbf{r}_2)}{|\mathbf{r}_1-\mathbf{r}_2|} d\mathbf{r}_2,
\end{equation}
and minimizing only over the one-dimensional parameter $\lambda \in \mathbb{R}$. The optimal value is
\begin{equation}
    \lambda = -\frac{(\mathbf{j}_{\mathrm{S}}|\boldsymbol{\alpha})}{\mu^{-1} (\boldsymbol{\beta}|\boldsymbol{\beta}) + (\rho|\alpha^2)} = -\frac{\coulpair{\mathbf{j}_{\mathrm{S}}}{\mathbf{j}_{\mathrm{S}}}}{\mu^{-1} \coulpair{\mathbf{j}_{\mathrm{S}}}{\mathbf{j}_{\mathrm{S}}} + 4\pi (\rho|\alpha^2)}
\end{equation}
and the resulting energy bound is
\begin{equation}
 \EMS(v,\mathbf{A}) \leq \Estd(v,\mathbf{A}) - \frac{\mu}{8\pi} \cdot \frac{\coulpair{\mathbf{j}_{\mathrm{S}}}{\mathbf{j}_{\mathrm{S}}}^2}{\coulpair{\mathbf{j}_{\mathrm{S}}}{\mathbf{j}_{\mathrm{S}}} + 4\pi \mu (\rho|\alpha^2)} 
\end{equation}
When $(\rho|\alpha^2)$ is finite and $\mu$ is sufficiently small,
\begin{equation}
 \EMS(v,\mathbf{A}) \lesssim \Estd(v,\mathbf{A}) - \frac{\mu}{8\pi} \cdot \coulpair{\mathbf{j}_{\mathrm{S}}}{\mathbf{j}_{\mathrm{S}}}.
\end{equation}
Hence, the Maxwell--Schr\"odinger energy is  bounded by the standard energy $\Estd(v,\mathbf{A})$ corrected by an attractive current-current interaction that resembles the relativistic Gaunt interaction~\cite{DYALL07}.

\subsection{Constrained-search formulation of MDFT}
\label{secMDFT_CS}

A constrained-search formulation is obtained by introducing the total vector potential as a basic variational parameter. A reparametrization $\boldsymbol{\alpha}' = \boldsymbol{\alpha} + \mathbf{A}$ leads to
\begin{equation}
 \label{eqEMSreparam}
 \begin{split}
  \EMS(v,\mathbf{A}) & = \inf_{\substack{\Gamma \\ \boldsymbol{\alpha}' \in \Aspace}} \Big( \frac{1}{2 \mu} (\boldsymbol{\beta}'-\mathbf{B}|\boldsymbol{\beta}'-\mathbf{B}) + \trace{ \Gamma (\op{T}_{\boldsymbol{\alpha}'} + \op{v} + \op{W}) }  \Big).
 \end{split}
\end{equation}
Note that it is here crucial that both the internal and external vector potential are in $\Aspace$, since this guarantees that the reparametrized minimization over $\boldsymbol{\alpha}'$ can be performed over the same domain. By contrast, if $\mathbf{A} \notin \Aspace$, then the minimization domain for $\boldsymbol{\alpha}'$ would be dependent on $\mathbf{A}$, ruining the universality of the theory. Note also that the minimization over $\Gamma$ is performed over all mixed states formed from pure states in $H^1_{\boldsymbol{\alpha}'}$. Since $\boldsymbol{\alpha}'$ is a variational parameter, and minimization is performed over all possible $\boldsymbol{\alpha}' \in \Aspace$, this does not prevent the construction of a universal density functional. Introducing a nested minimization, one may identify the functional $\FGH$ from Eq.~\eqref{eqFGHDef} above,
\begin{equation}
 \begin{split}
   \EMS(v,\mathbf{A}) & = \inf_{\substack{\rho \in X \\ \boldsymbol{\alpha}' \in \Aspace}} \Big( (\rho|v) + \frac{1}{2 \mu} (\boldsymbol{\beta}'-\mathbf{B}|\boldsymbol{\beta}'-\mathbf{B}) + \inf_{\Gamma \mapsto \rho}  \trace{ \Gamma (\op{T}_{\boldsymbol{\alpha}'} + \op{W}) }  \Big)
              \\
  & = \frac{1}{2 \mu} (\mathbf{B}|\mathbf{B}) + \inf_{\substack{\rho \in X \\ \boldsymbol{\alpha}' \in \Aspace}} \Big( (\rho|v) - \frac{1}{\mu} (\boldsymbol{\beta}'|\mathbf{B}) + \frac{1}{2 \mu} (\boldsymbol{\beta}'|\boldsymbol{\beta}') + \FGH(\rho,\boldsymbol{\alpha}')  \Big)
 \end{split}
\end{equation}
Defining the universal functional
\begin{equation}
 \begin{split}
   \label{eqfMSdef}
   \fMS(\rho,\boldsymbol{\alpha}') = \frac{1}{2 \mu} (\boldsymbol{\beta}'|\boldsymbol{\beta}') + \FGH(\rho,\boldsymbol{\alpha}')
 \end{split}
\end{equation}
now allows one to write
\begin{equation}
 \begin{split}
    \label{eqEMSfromfMS}
   \EMS(v,\mathbf{A})  & = \frac{1}{2 \mu} (\mathbf{B}|\mathbf{B}) + \inf_{\substack{\rho \in X \\ \boldsymbol{\alpha}' \in \Aspace}} \Big( (\rho|v) - \frac{1}{\mu} (\boldsymbol{\beta}'|\mathbf{B}) + \fMS(\rho,\boldsymbol{\alpha}')  \Big).
 \end{split}
\end{equation}
Here, the external potentials $(v,\mathbf{A})$ have been completely isolated from the remaining, universal functional $\fMS$. Moreover, the ground state energy is obtained from a straightforward variational principle involving global minimization over $(\rho,\boldsymbol{\alpha}')$. 

Two different semi-universal constrained-search functionals can also be defined. Nesting the minimizations so that minimization over $\rho$ is performed after $\boldsymbol{\alpha}'$ leads to the functional
\begin{equation}
  \label{eqFMSdef}
 \begin{split}
  \FMS(\rho,\mathbf{A}) & = \frac{1}{2 \mu} (\boldsymbol{\beta}'-\mathbf{B}|\boldsymbol{\beta}'-\mathbf{B}) + \inf_{\Gamma \mapsto \rho}  \trace{ \Gamma (\op{T}_{\boldsymbol{\alpha}'} + \op{W}) }  
             \\
 & = \frac{1}{2\mu} (\mathbf{B}|\mathbf{B}) + \inf_{\boldsymbol{\alpha}' \in \Aspace} \Big(-\frac{1}{\mu} (\boldsymbol{\beta}'|\mathbf{B}) + \fMS(\rho,\boldsymbol{\alpha}')  \Big),
 \end{split}
\end{equation}
while the opposite nesting yields
\begin{equation}
 \begin{split}
   \eMS(v,\boldsymbol{\alpha}') & = \frac{1}{2 \mu} (\boldsymbol{\beta}'|\boldsymbol{\beta}') + \inf_{\Gamma}  \trace{ \Gamma (\op{T}_{\boldsymbol{\alpha}'} + \op{v} + \op{W}) }  \Big)
              \\
  & = \inf_{\rho \in X} \Big( (\rho|v) + \fMS(\rho,\boldsymbol{\alpha}')  \Big).
 \end{split}
\end{equation}
It then follows that the ground-state energy can be obtained from two different variational principles,
\begin{equation}
 \begin{split}
   \label{eqEMSfromFMSandeMS}
   \EMS(v,\mathbf{A}) & = \inf_{\rho \in X} \Big( (\rho|v) + \FMS(\rho,\mathbf{A})\Big)
              \\
      & = \frac{(\mathbf{B}|\mathbf{B})}{2\mu} + \inf_{\boldsymbol{\alpha}'\in\Aspace} \Big(-\frac{(\boldsymbol{\beta}'|\mathbf{B})}{\mu} + \eMS(v,\boldsymbol{\alpha}') \Big).
 \end{split}
\end{equation}

In the ground state energy functional $\EMS(v,\mathbf{A})$, and the three constrained-search functionals $\fMS(\rho,\boldsymbol{\alpha}')$, $\eMS(\rho,\boldsymbol{\alpha}')$, and $\FMS(\rho,\mathbf{A})$ defined above, the second argument is a magnetic vector potential. By exploiting gauge invariance together with the fact that $\boldsymbol{\kappa}'$ and $\boldsymbol{\beta}'$ are little more than alternative representations of $\boldsymbol{\alpha}'$ (and similarly for $\mathbf{J},\mathbf{B},\mathbf{A}$), it is easy to reformulate the functionals so that the second argument is a magnetic field or a current density instead. For definiteness, we define
\begin{align}
   \FMSb(\rho,\mathbf{b}) & = \inf_{\substack{\mathbf{a} \in \Aspace \\ \nabla\times\mathbf{a} = \mathbf{b}}} \FMS(\rho,\mathbf{a}),
             \\
   \FMSj(\rho,\mathbf{j}) & = \inf_{\substack{\mathbf{a} \in \Aspace \\ \nabla\times\nabla\times\mathbf{a} = -\mu\mathbf{j}}} \FMS(\rho,\mathbf{a}),
\end{align}
and similarly for $\EMS$, $\eMS$ and $\fMS$. It should be noted, however, that these minimizations are trivial, since the choice of function spaces actually allows the relation $-\mu \mathbf{j} = \nabla\times\mathbf{b}$ to be inverted. By gauge invariance, the remaining non-invertibility of $\mathbf{b} = \nabla\times\mathbf{a}$ does not matter. Specifically,
\begin{align}
   \FMSb(\rho,\mathbf{B}) & = \FMS(\rho,\mathbf{A} + \nabla \chi), \quad \mathbf{B} = \nabla\times\mathbf{A},
\end{align}
for all $\chi$ such that $\nabla\chi \in \Aspace$. 

By abuse of notation, the same symbol will be used for all three versions of the underlying functional---e.g., $\FMS(\rho,\mathbf{A})$, $\FMSb(\rho,\mathbf{B})$, and $\FMSj(\rho,\mathbf{J})$ will be written $\FMS(\rho,\mathbf{A})$, $\FMS(\rho,\mathbf{B})$, and $\FMS(\rho,\mathbf{J})$. With this notational convention Eq.~\eqref{eqFMSdef} can alternatively be expressed as
\begin{align}
  \FMS(\rho,\mathbf{B}) & = \frac{1}{2\mu} (\mathbf{B}|\mathbf{B}) + \inf_{\boldsymbol{\beta}' \in \Bdiv} \Big(-\frac{1}{\mu} (\boldsymbol{\beta}'|\mathbf{B}) + \fMS(\rho,\boldsymbol{\beta}')  \Big),
          \\
  \FMS(\rho,\mathbf{A}) & = \frac{1}{2\mu} (\mathbf{B}|\mathbf{B}) + \inf_{\boldsymbol{\kappa}' \in \Jdiv} \Big((\boldsymbol{\kappa}'|\mathbf{A}) + \fMS(\rho,\boldsymbol{\kappa}')  \Big),
          \\
  \FMS(\rho,\mathbf{J}) & = \frac{\mu}{8\pi} \coulpair{\mathbf{J}}{\mathbf{J}} + \inf_{\boldsymbol{\alpha}' \in \Aspace} \Big((\boldsymbol{\alpha}'|\mathbf{J}) + \fMS(\rho,\boldsymbol{\alpha}')  \Big).
\end{align}

\subsection{Reformulation using convex analysis I: generalized Lieb formulation}
\label{secMDFT_LIEB}

The ground state functional and the constrained-search functionals defined in the previous section have non-trivial mathematical properties, which can be characterized in terms of the convex analysis concepts reviewed in Sec.~\ref{secCVXANL}.

We first turn turn to the convexity properties of the defined functionals as well as their interrelations through Legendre--Fenchel transformations. In order to distinguish partial Legendre--Fenchel transformations affecting the first and second argument, respectively, we now modify the notation slightly. Transformation with respect to the first (second) argument will be denoted by $\wedge$ and $\vee$ ($\bigtriangleup$ and $\bigtriangledown$) superscripts. Moreover, the pairing term in the transformation w.r.t.\ the second argument will be taken to be $\tfrac{1}{\mu} (\boldsymbol{\beta}'|\mathbf{B})$ rather than $-(\boldsymbol{\beta}'|\mathbf{B})$. With these conventions, one may identify the Legendre--Fenchel transform
\begin{align}
   \EMS(v,\mathbf{B}) & = \FMS^{\wedge}(v,\mathbf{B}) = \inf_{\rho\in X} \big( (\rho|v) + \FMS(\rho,\mathbf{B}) \big),
\end{align}
in the first part of Eq.~\eqref{eqEMSfromFMSandeMS}. Lieb's original proof of convexity and l.s.c. for the universal Levy--Lieb functional in standard DFT requires only trivial adaptations to apply also to $\FMS(\rho,\mathbf{B})$ as a functional of $\rho$ with $\mathbf{B}$ fixed. Hence, the above transformation can be inverted,
\begin{align}
   \FMS(v,\mathbf{B}) = \EMS^{\vee}(\rho,\mathbf{B}) = \inf_{v\in X^*} \big( \EMS(v,\mathbf{B}) - (\rho|v) \big).
\end{align}

Both the above functionals are paraconcave in $\mathbf{B}$. In order to exploit this fact, it is useful to shift the ground state energy by the magnetic-field energy $g_{\mu}(\mathbf{B}) = \frac{1}{2\mu} \|\mathbf{B}\|^2$,
\begin{align}
   \EMScvx(v,\mathbf{B}) = \EMS(v,\mathbf{B}) - \frac{1}{2\mu} \|\mathbf{B}\|^2,
\end{align}
yielding a functional that is jointly concave in $(v,\mathbf{B})$. This can be seen by noting that Eqs.~\eqref{eqEMSreparam} and \eqref{eqEMSfromfMS} are instances of the general minimization over affine functions in Eq.~\eqref{eqMINAFFPARAM}. The concavity properties of $\EMScvx$ are not restricted to the $(v,\mathbf{B})$-representation: $\EMScvx(v,\mathbf{A})$ is jointly concave in $(v,\mathbf{A})$ and $\EMScvx(v,\mathbf{J})$ is jointly concave in $(v,\mathbf{J})$.

Shifting the half-universal functional $\FMS$ by the magnetic field energy yields the convex-concave functional
\begin{align}
   \FMScvx(\rho,\mathbf{B}) = \FMS(\rho,\mathbf{B}) - \frac{1}{2\mu} \|\mathbf{B}\|^2,
\end{align}
that is, $\FMScvx$ is convex in $\rho$ and concave in $\mathbf{B}$. Once again, the convexity properties are not specific to the $(\rho,\mathbf{B})$-representation---$\FMScvx(\rho,\mathbf{A})$ and $\FMScvx(\rho,\mathbf{J})$ are convex-concave too.

By inspection of Eqs.~\eqref{eqEMSfromfMS} and \eqref{eqEMSfromFMSandeMS}, it is seen that the shifted ground-state energy can be obtained as a full Legendre--Fenchel transform of the universal functional, $\EMScvx = \fMS^{\wedge\bigtriangleup}$, and also as half-transforms of two semi-universal functionals, $\EMScvx = \FMScvx^{\wedge} = \eMS^{\bigtriangleup}$. In detail,
\begin{align}
          \label{eqEMScvxfromfMS}
   \EMScvx(v,\mathbf{B}) & = \fMS^{\wedge\bigtriangleup}(v,\mathbf{B}) = \inf_{\rho,\boldsymbol{\beta}'} \big( (\rho|v) - \frac{1}{\mu} (\boldsymbol{\beta}'|\mathbf{B}) + \fMS(\rho,\boldsymbol{\beta}') \big)
              \\
         & = \FMScvx^{\wedge}(v,\mathbf{B}) = \inf_{\rho,\boldsymbol{\beta}'} \big( (\rho|v) + \FMScvx(\rho,\mathbf{B}) \big)
               \\
          & = \eMS^{\bigtriangleup}(v,\mathbf{B}) = \inf_{\boldsymbol{\beta}'} \big( -\frac{1}{\mu} (\boldsymbol{\beta}'|\mathbf{B}) + \eMS(v,\boldsymbol{\beta}') \big).
        \label{eqEMScvxfromeMS}
\end{align}

Also the semi-universal functional $\FMScvx$ has multiple representations as a Legendre--Fenchel transformation. It is a full Legendre--Fenchel transform of the other semi-universal functional, $\FMScvx = \eMS^{\bigtriangleup\vee}$, and half-transforms of the shifted ground state energy and universal functionals, $\FMScvx = \EMScvx^{\vee} = \fMS^{\bigtriangleup}$. In detail, the half-transforms are
\begin{align}
         \label{eqFMScvxfromEMScvx}
  \FMScvx(\rho,\mathbf{B}) & = \EMScvx^{\vee}(\rho,\mathbf{B}) = \sup_{v\in X^*} \big( \EMScvx(\rho,\mathbf{A}) - (\rho|v) \big)
              \\
       & = \fMS^{\bigtriangleup}(\rho,\mathbf{B}) = \inf_{\boldsymbol{\beta}' \in \Bdiv} \big( -\frac{1}{\mu} (\boldsymbol{\beta}'|\mathbf{B}) + \fMS(\rho,\boldsymbol{\beta}') \big).
         \label{eqFMScvxfromfMS}
\end{align}
Inserting Eq.~\eqref{eqEMScvxfromeMS} into Eq.~\eqref{eqFMScvxfromEMScvx} yields the full transformation,
\begin{align}
   \label{eqFMScvxfromeMS}
   \FMScvx(\rho,\mathbf{B}) & = \eMS^{\bigtriangleup \vee}(\rho,\mathbf{B}) = \sup_{v \in X^*} \inf_{\boldsymbol{\beta}' \in \Bdiv} \big( -(\rho|v) - \frac{1}{\mu} (\boldsymbol{\beta}'|\mathbf{B}) + \eMS(v,\boldsymbol{\beta}') \big).
\end{align}
An alternative expression with the maximization and minimization interchanged is obtained by inserting Eq.~\eqref{eqfMSfromeMS} into Eq.~\eqref{eqFMScvxfromfMS},
\begin{align}
   \label{eqFMScvxfromeMS2}
   \FMScvx(\rho,\mathbf{B}) & = \eMS^{\vee \bigtriangleup}(\rho,\mathbf{B}) = \inf_{\boldsymbol{\beta}' \in \Bdiv} \sup_{v \in X^*} \big( -(\rho|v) - \frac{1}{\mu} (\boldsymbol{\beta}'|\mathbf{B}) + \eMS(v,\boldsymbol{\beta}') \big).
\end{align}

\begin{figure}[tb]
 \caption{\label{figMDFTrel} Graphical representation of the relations between functionals in the convex formulation of MDFT. Double arrows represent Legendre--Fenchel transformations, while single arrows represent MY regularization and additive shifts of the functionals.}
 \includegraphics[width=0.8\textwidth]{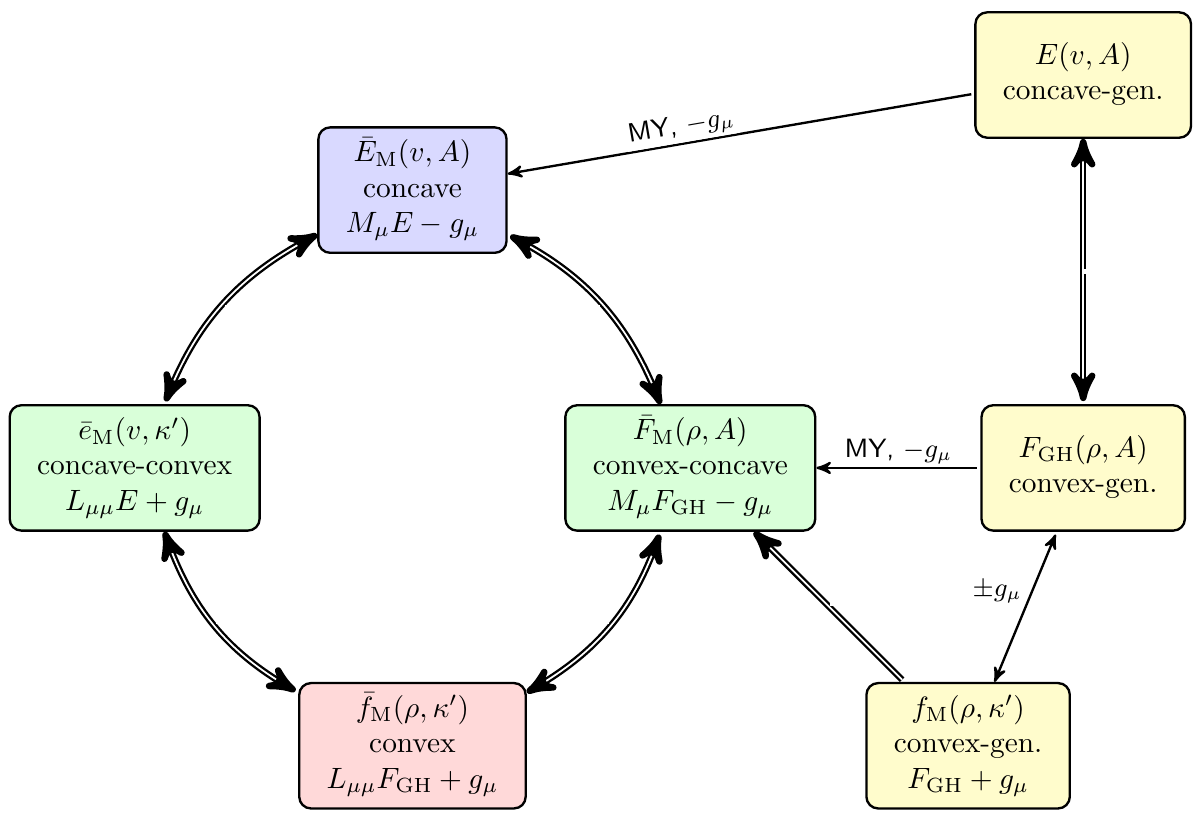}
\end{figure}

Turning next to the properties of $\fMS$, we rewrite Eq.~\eqref{eqfMSdef} into
\begin{equation}
 \begin{split}
   \fMS(\rho,\boldsymbol{\alpha}') = \frac{1}{2 \mu} (\boldsymbol{\beta}'|\boldsymbol{\beta}') + \FGHcvx(\rho,\boldsymbol{\alpha}') + \frac{1}{2} (\rho|\alpha'^2),
 \end{split}
\end{equation}
where we temporarily switch back to regarding $\fMS$ as a functional of $\boldsymbol{\alpha}'$ since the last two terms are separately gauge dependent. The sum of the terms, however, is gauge invariant. Each term above is separately convex in $\rho$, so that the sum is also convex. The first and third terms are also convex in $\boldsymbol{\alpha}'$. The second term, however, is concave in $\boldsymbol{\alpha}'$, leaving the concavity properties of the functional unclear. Without a proof of joint convexity and l.s.c.\ of $\fMS$, it is necessary to distinguish this functional from its double Legendre--Fenchel transform. The functionals $\fMS$ and $\eMS$ do, however, constitute a transform pair w.r.t.\ the first argument,
\begin{equation}
   \eMS(v,\boldsymbol{\beta}') = \fMS^{\wedge}(v,\boldsymbol{\beta}') = \inf_{\rho \in X} \big( (\rho|v) +  \fMS(\rho,\boldsymbol{\beta}') \big)
\end{equation}
and
\begin{equation}
   \label{eqfMSfromeMS}
   \fMS(\rho,\boldsymbol{\beta}') = \eMS^{\vee}(\rho,\boldsymbol{\beta}') = \sup_{v \in X^*} \big( \eMS(v,\boldsymbol{\beta}') - (\rho|v) \big).
\end{equation}
In order to obtain a theory with invertible Legendre--Fenchel transformations, it is useful to define the semi-universal functional
\begin{equation}
   \label{eqeMScvxfromEMScvx}
   \eMScvx(v,\boldsymbol{\beta}') = \EMScvx^{\bigtriangledown}(v,\boldsymbol{\beta}') = \sup_{\mathbf{B} \in \Bdiv} \big( \EMScvx(v,\mathbf{B}) + \frac{1}{\mu} (\boldsymbol{\beta}'|\mathbf{B}) \big)
\end{equation}
and the universal functional
\begin{equation}
   \label{eqfMScvxdef}
   \fMScvx(\rho,\boldsymbol{\beta}') = \EMScvx^{\vee\bigtriangledown}(\rho,\boldsymbol{\beta}') = \sup_{\substack{v\in X^* \\ \mathbf{B} \in \Bdiv}} \big( \EMScvx(v,\mathbf{B}) - (\rho|v) + \frac{1}{\mu} (\boldsymbol{\beta}'|\mathbf{B}) \big).
\end{equation}
All of the expressions from Eq.~(\ref{eqEMScvxfromfMS}--\ref{eqFMScvxfromeMS2}) hold with $\eMS$ and $\fMS$ replaced $\eMScvx$ and $\fMScvx$, respectively. In particular, one obtains the transform pair $\EMScvx = \fMScvx^{\wedge\bigtriangleup}$ and $\fMScvx = \EMScvx^{\vee\bigtriangledown}$. The cycle of partial Legendre--Fenchel transformations is summarized in Fig.~\ref{figMDFTrel}. (As a technical remark, the partially transformed functionals $\eMScvx(v,\boldsymbol{\beta}')$ and $\FMScvx(\rho,\mathbf{B})$ are saddle-functionals that satisfy dual minimax and maximin principles~\cite{ROCKAFELLAR_PJM25_597}. The general theory of saddle functionals does not guarantee that these are unique saddle functionals, only that they are two equivalence classes of functionals that can be used interchangeably in the variational principles. The equalities $\eMScvx = \FMScvx^{\vee \bigtriangleup} = \FMScvx^{\bigtriangleup\vee}$ and $\FMScvx = \eMScvx^{\wedge \bigtriangledown} = \eMScvx^{\bigtriangledown\wedge}$ should thus rigorously be understood as statements about equivalence classes.)

The above formulation includes an explicit variational principle (in Eq.~\eqref{eqfMScvxdef}) for the universal functional $\fMScvx$ that generalizes Lieb's variational principle in standard DFT. This variational principle can be slightly generalized and used to provide a practical framework for computing systematic approximations to Adiabatic Connection curves that characterize the exact functional. Several studies have employed this method in the standard DFT framework~\cite{COLONNA_JCP110_2828,FRYDEL_JCP112_5292,SAVIN_JCP115_6827,WU_JCP118_2498,TEALE_JCP130_104111,TEALE_JCP132_164115,TEALE_JCP133_164112,STROMSHEIM_JCP135_194109} and the BDFT framework~\cite{REIMANN_JCTC13_4089}. This generalization is done by replacing the electron--electron repulsion $\op{W}$ in Eq.~\eqref{eqEMSdef} by $\lambda \op{W}$, where $\lambda$ is a scaling factor.

\subsection{Reformulation using convex analysis II: MY and LL regularization}
\label{secMDFT_MY}

By inspection of Eq.~\eqref{eqEMSreparam}, the Maxwell--Schr\"odinger ground state functional is seen to be the infimal convolution w.r.t.\ the second argument of the Schr\"odinger functional and the magnetic self-energy $g_{\mu}(\mathbf{B}) = \|\mathbf{B}\|^2 / 2\mu = (\mathbf{B}|\mathbf{B})/2\mu$.
\begin{equation}
  \label{eqEMSasInfConv}
  \EMS(v,\mathbf{B}) = (\Estd \Box g_{\mu})(v,\mathbf{B}) = \inf_{\boldsymbol{\beta}'\in\Aspace} \big( \Estd(v,\boldsymbol{\beta}') + \frac{(\mathbf{B}-\boldsymbol{\beta}'|\mathbf{B}-\boldsymbol{\beta}')}{2\mu} \big),
\end{equation}
where we  let $\Box$ denote convolution w.r.t.\ the second argument. We have here also chosen to regard the ground state energy as a functional of $(v,\mathbf{B})$, since $\mathbf{B}$ and $\boldsymbol{\beta}'$ are elements of the Hilbert space $\Bdiv$. Due to the specific form of $g_{\mu}$, the functional $\EMS$ is a Moreau--Yosida regularization of $\Estd$. Previous work has considered MY regularization of the scalar potential in standard DFT~\cite{KVAAL_JCP140_18A518}, which corresponds to the MY regularization of the first argument (i.e., $v$) in the present setting. Due to concavity in $v$, this results in improved regularity so that the functional derivative, in addition to the sub- or supergradient, is well-defined~\cite{KVAAL_JCP140_18A518}. However, the present situation is more involved since $\Estd(v,\mathbf{A})$ is neither concave nor convex in $\mathbf{A}$. The MY regularized functional $\EMS = \Estd\Box g_{\mu}$ then does not automatically have improved regularity.

From inspection of Eq.~\eqref{eqFMSdef}, one finds
\begin{equation}
  \FMS(\rho,\mathbf{B}) = (\FGH \Box g_{\mu})(\rho,\mathbf{B}).
\end{equation}
Consequently, the functionals shifted by the magnetic field energy can be written
\begin{align}
  \EMScvx(\rho,\mathbf{B}) = (\Estd \Box g_{\mu})(\rho,\mathbf{B}) - g_{\mu}(\mathbf{B}),
     \\
  \FMScvx(\rho,\mathbf{B}) = (\FGH \Box g_{\mu})(\rho,\mathbf{B}) - g_{\mu}(\mathbf{B}).
\end{align}
For comparison, the constrained-search functionals $\eMS$ and $\fMS$, are simply shifted by $g_{\mu}$ rather than being MY regularizations,
\begin{align}
     \eMS(v,\boldsymbol{\alpha}') = \Estd(v,\boldsymbol{\alpha}') + g_{\mu}(\boldsymbol{\alpha}'),
               \\
     \fMS(\rho,\boldsymbol{\alpha}') = \FGH(\rho,\boldsymbol{\alpha}') + g_{\mu}(\boldsymbol{\alpha}').
\end{align}
The MY regularizations and additive shifts that relate the above functionals to the standard energy functional and the Grayce--Harris functional are illustrated in Fig.~\ref{figMDFTrel}.

Inserting the identity
\begin{equation}
   \frac{1}{\mu} (\boldsymbol{\beta}'|\mathbf{B}) = g_{\mu}(\mathbf{B}) - g(\mathbf{B}-\boldsymbol{\beta}') + g_{\mu}(\boldsymbol{\beta}')
\end{equation}
into Eq.~\eqref{eqeMScvxfromEMScvx} above yields
\begin{equation}
 \begin{split}
   \eMScvx(v,\boldsymbol{\beta}') = \sup_{\mathbf{B} \in \Bdiv} \big( \EMS(v,\mathbf{B}) - g(\mathbf{B}-\boldsymbol{\beta}') + g_{\mu}(\boldsymbol{\beta}') \big) = -(g_{\mu} \Box -\EMS)(v,\boldsymbol{\beta}') + g_{\mu}(\boldsymbol{\beta}').
 \end{split}
\end{equation}
Finally, inserting Eq.~\eqref{eqEMSasInfConv} yields
\begin{equation}
 \begin{split}
   \eMScvx(v,\boldsymbol{\beta}') = -(g_{\mu} \Box -(g_{\mu} \Box \Estd))(v,\boldsymbol{\beta}') + g_{\mu}(\boldsymbol{\beta}') = L_{\mu\mu} E(v,\boldsymbol{\beta}') + g_{\mu}(\boldsymbol{\beta}').
 \end{split}
\end{equation}
Very similar reasoning shows that
\begin{equation}
 \begin{split}
   \fMScvx(\rho,\boldsymbol{\beta}') = -(g_{\mu} \Box -(g_{\mu} \Box \FGH))(\rho,\boldsymbol{\beta}') + g_{\mu}(\boldsymbol{\beta}') = L_{\mu\mu} \FGH(\rho,\boldsymbol{\beta}') + g_{\mu}(\boldsymbol{\beta}').
 \end{split}
\end{equation}
Comparing to the generic Eq.~\eqref{eqLLREGDEF}, the above two expressions can be recognized as the proximal hull, or a LL regularization, of the Schr\"odinger functional $\Estd$ and the semi-universal functional $\FGH$.

\subsection{Reformulation using convex analysis III: differentiable LL regularization}
\label{secMDFT_DIFF}

In the above treatment, the parameter $\mu$ plays a dual role as it is used both to obtain $\EMS$ from $E$ and in the cycle of Legendre--Fenchel transformations summarized in Fig.~\ref{figMDFTrel}. In fact, at the cost of somewhat complicating the relationship to the constrained-search functionals, it is possible to operate with two different regularization parameters, $\zeta$ and $\lambda$. The parameter $\zeta$ is used in the MY regularization of the Schr\"odinger energy and Grayce--Harris functional: $\EMS(v,\mathbf{B}) = M_{\zeta} \Estd(v,\mathbf{B})$ and $\FMS(\rho,\mathbf{B}) = M_{\zeta} \FGH(\rho,\mathbf{B})$. The parameter $\lambda \leq \zeta$ is used to define the jointly concave functional
\begin{equation}
   \EMScvx(v,\mathbf{B}) = -g_{\lambda}(\mathbf{B}) + M_{\zeta} \Estd(v,\mathbf{B}) = -g_{\zeta}(\mathbf{B}) + M_{\zeta} \Estd(v,\mathbf{B}) + \big(g_{\zeta}(\mathbf{B}) - g_{\lambda}(\mathbf{B})\big),
\end{equation}
where $-g_{\zeta} + M_{\zeta} \Estd$ and $g_{\zeta} - g_{\lambda}$ are both concave, and the convex-concave functional
\begin{equation}
    \FMScvx(\rho,\mathbf{B}) = -g_{\lambda}(\mathbf{B}) + M_{\zeta} \FGH(\rho,\mathbf{B}).
\end{equation}
The parameter $\lambda$ is also used in the cycle of Legendre--Fenchel transforms,
\begin{align}
  \eMScvx(v,\boldsymbol{\beta}') & = \EMScvx^{\bigtriangledown}(v,\boldsymbol{\beta}') = \sup_{\mathbf{B} \in \Bdiv} \big( \EMScvx(v,\mathbf{B}) + \frac{1}{\lambda} (\boldsymbol{\beta}'|\mathbf{B}) \big),
              \\
  \fMScvx(\rho,\boldsymbol{\beta}') & = \FMScvx^{\bigtriangledown}(\rho,\boldsymbol{\beta}') = \sup_{\mathbf{B} \in \Bdiv} \big( \FMScvx(\rho,\mathbf{B}) + \frac{1}{\lambda} (\boldsymbol{\beta}'|\mathbf{B}) \big),
\end{align}
and similarly for the inverse partial transformations. The above expressions are equivalent to the following LL regularizations:
\begin{align}
  \eMScvx(v,\boldsymbol{\beta}') & = \frac{\|\boldsymbol{\beta}'\|^2}{2\lambda} + L_{\lambda\zeta} \Estd(v,\boldsymbol{\beta}'),
              \\
  \fMScvx(\rho,\boldsymbol{\beta}') & = \frac{\|\boldsymbol{\beta}'\|^2}{2\lambda} + L_{\lambda \zeta} \FGH(\rho,\boldsymbol{\beta}').
\end{align}
Equivalence follows by writing out definition of the LL regularizations and expanding terms representing the norm of magnetic field differences.

Besides the additional flexibility, the advantage of this generalization is that functional differentiability can be established. Differentiability with respect to the first argument (scalar potentials and densities) has been shown to be a non-trivial issue and the standard DFT functionals are in fact not differentiable in the ordinary sense~\cite{LAMMERT_IJQC107_1943}. For a fixed $v$, it follows from Eq.~\eqref{eqPARAMAGBOUND} that the Schr\"odinger energy functional is bounded by a quadratic function, $E(v,\mathbf{B}) \geq -\mathrm{const}\cdot(\|\mathbf{B}\|^2+1)$. For such quadratically bounded functions, Theorem~4.1 due to Attouch and Aze is applicable~\cite{ATTOUCH_AIHP11_289}.  When $\lambda < \zeta$, this theorem guarantees that $L_{\lambda\zeta} E$, and therefore also $\eMScvx$, has a H\"older continuous (Fr\'echet) derivative as a function of the magnetic field (for a fixed $v$). Because the constraint $\Gamma \mapsto \rho$ only raises the energy, the bound in Eq.~\eqref{eqPARAMAGBOUND} can be straightforwardly adapted to a quadratic bound on the Grayce--Harris functional,
\begin{equation}
   (\rho|v) + \FGH(\rho,\mathbf{B}) = -\frac{\|\mathbf{B}\|^2}{2\zeta} + \inf_{\Gamma \mapsto \rho} \Big( \frac{\|\mathbf{B}\|^2}{2\zeta} + \trace{\Gamma (\op{T}_{\mathbf{A}} + v + W)} \Big) \geq -\frac{\|\mathbf{B}\|^2}{2\zeta} + \EMS(v,0),
\end{equation}
where $\EMS = M_{\zeta} \Estd$. Hence, for $\lambda < \zeta$ and a fixed $\rho$, it follows that also $\fMScvx = g_{\lambda} + L_{\lambda\zeta} \FGH$ is differentiable as a function of the magnetic field.

\subsection{Current densities, subgradients, and a Hohenberg--Kohn-like mapping}
\label{secMDFT_HK}

Two obstacles that to date have prevented the formulation of a CDFT based on the physical current density get natural resolutions in the present formulation. Firstly, the physical current density in Eq.~\eqref{eqJphysDef} explicitly depends on the external potential, making it difficult to construct a universal functional of $(\rho,\mathbf{j})$. Secondly, it has been shown that a physical CDFT is inconsistent with the standard variational principle~\cite{LAESTADIUS_PRA91_032508}. It is implicit in this result that the wave function (or density matrix) is the only variational parameter and that physical current densities are obtained in the standard way from the wave function and the external magnetic vector potential. In the present Maxwell--Schr\"odinger formulation, the internal magnetic field is an additional variational parameter and the physical current density can be obtained as
\begin{equation}
  \mathbf{j}_{\boldsymbol{\alpha}'} = \frac{1}{2} \sum_k n_k \phi_k(\mathbf{r})^{\dagger} \boldsymbol{\sigma} (\boldsymbol{\sigma}\cdot\boldsymbol{\pi}_{\boldsymbol{\alpha}'}) \phi_k(\mathbf{r}) + \text{c.c.},
\end{equation} 
where $n_k$ are natural occupation numbers and $\phi_k$ natural spinors. Both obstacles are circumvented because the internal magnetic field, and its associated current density $\boldsymbol{\kappa} = -\nabla\times\boldsymbol{\beta}/\mu$, is a variational degree of freedom that is completely independent of the wave function. The current density $\mathbf{j}_{\boldsymbol{\alpha}'}$ depends directly on the wave function, but does not figure in the variational principle. Nonetheless, as will be detailed below, the resulting theory is physical CDFT because energy-minimization leads to the condition $\boldsymbol{\kappa} = \mathbf{j}_{\boldsymbol{\alpha}'}$. Hence, although $\boldsymbol{\kappa}$ is an independent parameter in the variational principle, its optimal value is always equivalent to the physical current density.

Consider the case when a minimizer $(\boldsymbol{\alpha},\psi)$ exists in Eq.~\eqref{eqEMSdef}. Further assume that this minimizer is unique up to gauge transformations. Then
\begin{equation}
   \EMS(v,\mathbf{A}) = \frac{1}{2\mu} (\boldsymbol{\beta}|\boldsymbol{\beta}) + \bra{\psi} \op{T}_{\boldsymbol{\alpha}+\mathbf{A}} + \op{v} + \op{W} \ket{\psi}.
\end{equation}
For any $\mathbf{a} \in \Aspace$, the Hellmann--Feynman theorem yields
\begin{equation}
   \frac{d}{d\epsilon} \EMS(v,\mathbf{A}+\epsilon \mathbf{a}) \Big|_{\epsilon=0} = (\mathbf{a}|\mathbf{j}).
\end{equation}
Any pure gauge term $\mathbf{a} = \nabla\chi$ yields a vanishing contribution, since $(\nabla\chi|\mathbf{j}) = -(\chi|\nabla\cdot\mathbf{j})$ and $\nabla\cdot\mathbf{j}=0$ for an energy-optimized state. After the reparametrization $\boldsymbol{\alpha}' = \boldsymbol{\alpha} + \mathbf{A}$, one may also write
\begin{equation}
   \EMS(v,\mathbf{A}) = \frac{1}{2\mu} (\mathbf{B}-\boldsymbol{\beta}'|\mathbf{B}-\boldsymbol{\beta}') + \bra{\psi} \op{T}_{\boldsymbol{\alpha}'} + \op{v} + \op{W} \ket{\psi},
\end{equation}
where $(\boldsymbol{\alpha}',\psi)$ is now the unique minimizer (up to gauge transformations) of the reparametrized minimization problem in Eq.~\eqref{eqEMSreparam}. The functional derivative now takes the form
\begin{equation}
   \frac{d}{d\epsilon} \EMS(v,\mathbf{A}+\epsilon \mathbf{a}) \Big|_{\epsilon=0} = \frac{1}{\mu} (\nabla\times\mathbf{a}|\mathbf{B}-\boldsymbol{\beta}') = -\frac{1}{\mu} (\mathbf{a}|\nabla\times\boldsymbol{\beta}).
\end{equation}
The above two forms are consistent, because any minimizer must satisfy Amp{\'e}re's circuital law, $\nabla\times\boldsymbol{\beta}=-\mu\mathbf{j}$. In the present setting, this equation is the stationarity condition for variations with respect to $\boldsymbol{\alpha}$. Without entering into technical details about functional derivatives, we may write informally
\begin{align}
  \frac{\delta \EMS(v,\mathbf{A})}{\delta v} & = \rho,
                \\
  \frac{\delta \EMS(v,\mathbf{A})}{\delta \mathbf{A}} & = \boldsymbol{\kappa} = \mathbf{j}.
\end{align}
For the concave functional $\EMScvx(v,\mathbf{A})$, one obtains 
\begin{align}
  \frac{\delta \EMScvx(v,\mathbf{A})}{\delta v} & = \rho,
                \\
  \frac{\delta \EMScvx(v,\mathbf{A})}{\delta \mathbf{A}} & = \boldsymbol{\kappa} + \mathbf{J} = \boldsymbol{\kappa}'.
\end{align}
and the alternative forms
\begin{align}
  \frac{\delta \EMScvx(v,\mathbf{B})}{\delta \mathbf{B}} & = -\frac{1}{\mu} (\boldsymbol{\beta} + \mathbf{B}) = -\frac{1}{\mu} \boldsymbol{\beta}',
                \\
  \frac{\delta \EMScvx(v,\mathbf{J})}{\delta \mathbf{J}} & = \boldsymbol{\alpha} + \mathbf{A} = \boldsymbol{\alpha}',
\end{align}
where gauge invariance of the last expression requires that $\boldsymbol{\alpha}' \in \Adiv$ is divergence free.

Let us now focus on the $(v,\mathbf{B})$-representation as it is straightforward to obtain analogous results in the $(v,\mathbf{A})$- and $(v,\mathbf{J})$-representations. Because $\EMScvx(v,\mathbf{B})$ is a concave functional, the above somewhat informal functional derivatives can be replaced by the rigorous notion of sub- and superdifferentials. For given external potentials $(v_0,\mathbf{B}_0)$, any minimizer $(\rho_0,\boldsymbol{\beta}'_0)$ in Eq.~\eqref{eqEMScvxfromfMS} is a supergradient. The functional value at an arbitrary external potential pair always lies below the linear extrapolation based on such a supergradient,
\begin{equation}
   \EMScvx(v,\mathbf{B}) \leq \EMScvx(v_0,\mathbf{B}_0) + (\rho_0|v-v_0) - \frac{1}{\mu} (\boldsymbol{\beta}'_0|\mathbf{B}-\mathbf{B}_0), \quad \forall v,\mathbf{B}.
\end{equation}
The minimizer $(\rho_0,\boldsymbol{\beta}'_0)$ need not be unique in order for this to hold.

Analogously, the functional $\fMScvx(\rho,\boldsymbol{\beta}')$ is convex. For a given density pair $(\rho_0,\boldsymbol{\beta}'_0)$, assume that a maximizer $(v_0,\mathbf{B}_0)$ exists in Eq.~\eqref{eqfMScvxdef}. Then the subgradient of $\fMScvx(v_0,\boldsymbol{\beta}'_0)$ is $(-v_0,-\mathbf{B}_0)$. The linear extrapolation based on the subgradient yields a global upper bound,
\begin{equation}
  \fMScvx(\rho,\boldsymbol{\beta}') \geq \fMScvx(\rho_0,\boldsymbol{\beta}'_0) - (\rho-\rho_0|v_0) + \frac{1}{\mu} (\boldsymbol{\beta}'-\boldsymbol{\beta}'_0|\mathbf{B}_0), \quad \forall \rho,\boldsymbol{\beta}'.
\end{equation}
Again, the maximizing potentials $(v_0,\mathbf{B}_0)$ need not be unique for this to hold.

The superdifferential $\overline{\partial} \EMScvx(v,\mathbf{B})$ is the set of all supergradients at $(v,\mathbf{B})$. This set can be empty, contain a single element, or contain multiple elements. Any minimizing pair $(\rho,\boldsymbol{\beta}')$ in Eq.~\eqref{eqEMScvxfromfMS} contributes an element $(\rho,\boldsymbol{\beta}')\in\overline{\partial} \EMScvx(v,\mathbf{B})$ to the superdifferential. Hence, more than element in $\overline{\partial} \EMScvx(v,\mathbf{B})$ implies a ground state degeneracy. When no minimizing pair exists, the superdifferential $\overline{\partial} \EMScvx(v,\mathbf{B})$ is the empty set.

The subdifferential $\underline{\partial} \fMScvx(\rho,\boldsymbol{\beta}')$ is the set of all subgradients at $(\rho,\boldsymbol{\beta}')$. For density pairs $(\rho,\boldsymbol{\beta}')$ that are not obtainable as the ground state density of any $(v,\mathbf{B})$, the subdifferential $\underline{\partial} \fMScvx(\rho,\boldsymbol{\beta}')$ is the empty set. When there a multiple potentials that all yield the density $(\rho,\boldsymbol{\beta}')$ as the minimizer, the subdifferential $\underline{\partial} \fMScvx(\rho,\boldsymbol{\beta}')$ has as many elements.

It is a general fact about convex functionals that there is a one-to-one mapping between non-empty subdifferentials and non-empty superdifferentials of the convex conjugate. In the present setting, there is a one-to-one mapping between subdifferentials $\underline{\partial} \fMScvx(\rho,\boldsymbol{\beta}')$ and superdifferentials $\overline{\partial} \EMScvx(v,\mathbf{B})$. This mapping can be expressed with the following equivalence:
\begin{equation}
   (\rho,\boldsymbol{\beta}') \in \overline{\partial} \EMScvx(v,\mathbf{B}) \iff (-v,-\mathbf{B}) \in \underline{\partial} \fMScvx(\rho,\boldsymbol{\beta}').
\end{equation}
Hence, the present theory admits a mapping that generalizes the usual Hohenberg--Kohn mapping in the standard DFT. This mapping exists for all positive values $\mu>0$ of the regularization parameter. In addition, the Maxwell--Schr\"odinger model tends to the standard Schr\"odinger model in the $\mu \to 0$ limit in the sense that
\begin{equation}
     \lim_{\mu \to 0^+} \EMS(v,\mathbf{A}) = \Estd(v,\mathbf{A})
\end{equation}
holds at least pointwise. We therefore speculate that the limit $\mu \to 0$ produces a Hohenberg--Kohn-like mapping between density pairs $(\rho,\mathbf{j})$ and potentials $(v,\mathbf{B})$ also in the standard Schr\"odinger model. This sheds new light on a long-standing open problem in the CDFT literature~\cite{DIENER_JP3_9417,PAN_IJQC110_2833,TELLGREN_PRA86_062506,LAESTADIUS_IJQC114_782}.

\section{Specialization to uniform magnetic fields}
\label{secUNIFORM}

Recent work has established a specialization of paramagnetic CDFT to
uniform magnetic fields~\cite{TELLGREN_1709_10400}. In this theory, the basic variables are the density $\rho$, the canonical
momentum $\mathbf{p} = \int \jpvec d\mathbf{r}$, and the paramagnetic
moment
$\mathbf{M}_{\mathrm{p}} = \int \mathbf{r}\times\jpvec d\mathbf{r} +
g_{\mathrm{e}} \mathbf{S}$. Due to the gauge-dependence and
non-extensivity of the paramagnetic moment, it is challenging to find
an approximate, practical Ansatz for the exchange-correlation
functional that preserves additivity over well-separated,
non-interacting subsystems~\cite{TELLGREN_1709_10400}.

In order to obtain a parallel specializiation of the above MDFT formulation to uniform magnetic fields, we first introduce a finite spatial domain $V_N$ and take the energy of a uniform field to be $\frac{1}{2\mu} \int_{V_N} B^2 d\mathbf{r} = |V_N| B^2 / 2\mu$. The $N$-dependence of the volume will be discussed shortly. Interpreting $\mu$ as a regularization parameter, rather than a physical parameter with a fixed empirical value, the volume $|V_N|$ can be made arbitrarily large while holding the ratio $\nu_N = \mu / |V_N|$ fixed. The energy can be written
\begin{equation}
  \EMS(v,\mathbf{B}) = \inf_{\substack{\rho \in X, \\ \boldsymbol{\alpha}'}} \Big( (\rho|v) + \frac{|\boldsymbol{\beta}'-\mathbf{B}|^2}{2\nu_N} + \inf_{\Gamma \mapsto \rho} \trace{\Gamma (\op{T}_{\boldsymbol{\alpha}'} + \op{W})} \Big),
\end{equation}
where the minimization is over all vector potentials $\boldsymbol{\alpha}'$ representing homogenous fields $\boldsymbol{\beta}'$. Taking the derivative with respect to $\mathbf{B}$ yields the physical magnetic moment,
\begin{equation}
  \mathbf{M} = 2 \frac{\partial \EMS(v,\mathbf{B})}{\partial \mathbf{B}} = 2 \, \frac{\mathbf{B}-\boldsymbol{\beta}'}{\nu_N} = -\frac{2}{\nu_N} \, \boldsymbol{\beta}.
\end{equation}
Hence, the internal magnetic field, $\boldsymbol{\beta} = - \nu_N \mathbf{M} / 2$, is essentially the magnetic moment in different units.

Unlike the paramagnetic moment $\mathbf{M}_{\mathrm{p}}$, the physical magnetic moment $\mathbf{M}$ is an extensive quantity. If $|V_N| = |V_1|$ is independent of $N$, also the internal magnetic field $\boldsymbol{\beta}$ becomes an extensive quantity. If the volume is instead scaled with the number of particles, $|V_N| = N |V_1|$, one obtains the more natural case that $\boldsymbol{\beta}$ is an intensive quantity. In what follows, we suppress the dependence on $N$ from the notation and write $\nu$.

The simplified setting offered by this reduction to a three-dimensional space of external and internal magnetic fields can be viewed as a theory in its own right. Alternatively, calculations in this setting can be viewed as idealized, schematic illustrations of the full theory. On this note, two model cases will be considered.

\subsection{Example: weak-field regime of an atomic system}

The weak-field regime of an atom with paramagnetic moment $2\gamma>0$ (directed anti-parallel to the magnetic field) and isotropic magnetizability $-2\xi < 0$ is described by
\begin{equation}
  \Estd(v,B) = \Estd(v,0) - \gamma |B| + \xi B^2,
\end{equation}
where $v(\mathbf{r}) = -Z/r$. Due to the even symmetry of this function it is sufficient to study the case $B,\beta' \geq 0$, leading to
\begin{equation}
    \EMS(v,B) = \inf_{\beta' \geq 0} \Big( \frac{(\beta'-B)^2}{2\nu} + \Estd(v,\beta') \Big)
      = \Estd(v,0) + \frac{\xi B^2 - \gamma B -\tfrac{1}{2} \nu \gamma^2}{1+2 \nu \xi}.
\end{equation}
The energy for $B \leq 0$ is obtained by mirroring the positive side, $\EMS(v,-B) = \EMS(v,B)$. From the above paraconcave functional of $B$ we can form the concave functional
\begin{equation}
    \EMScvx(v,B) = \EMS(v,B) - \frac{B^2}{2\nu}
 =  \Estd(v,0) - \frac{\tfrac{1}{2 \nu} B^2 + \gamma |B| + \tfrac{1}{2} \nu \gamma^2}{1+2 \nu \xi}
 =  \Estd(v,0) - \frac{\tfrac{1}{2 \nu} \big(|B| + \nu \gamma\big)^2}{1+2 \nu \xi}.
\end{equation}
A Legendre--Fenchel transform of the $B$ variable now yields
\begin{equation}
    \eMScvx(v,\beta') = \sup_{B} \big( \EMScvx(v,B) - \frac{\beta' B}{\nu} \big)
 =  \begin{cases}
        \Estd(v,0) - \frac{\nu}{2 (1+2\nu \xi)} \gamma^2, & \text{if\ } |\beta'| \leq \nu \gamma /(1+2\nu \xi), \\
        \Estd(v,0) - \gamma |\beta'| + (\xi + \frac{1}{2\nu}) \beta'^2, & \text{otherwise}.
   \end{cases}
\end{equation}
For comparison, the constrained-search functional is
\begin{equation}
   \eMS(v,\beta') = \Estd(v,0) - \gamma |\beta'| + \big(\xi + \frac{1}{2\nu} \big) \beta'^2.
\end{equation}
This functional is not convex, but agrees with the convex functional $\eMScvx$ on the subdomain $|\beta'| \geq \nu \gamma / (1+2\nu\xi)$. In the other part of the domain, $|\beta'| < \nu \gamma / (1+2\nu\xi)$, the constrained-search functional $\eMS$ is strictly larger than the convex envelope $\eMScvx$.

\begin{figure}[tb!]
\centering
\caption{\label{figWFATOM} Functionals related to the model energy for an atom in the weak-field regime. MY regularization of the non-convex $\Estd(v,b)$ (black dotted curve) results in the paraconcave functional $\EMS(v,b)$ (blue dash-dot curve), which is non-differentiable at $b=0$. Subtraction of the magnetic field energy yields the concave functional $\EMScvx(v,b)$ (solid blue curve). Addition of the magnetic field energy to $\Estd(v,b)$ yields the functional $\eMS(v,b)$ (red dashed curve), which remains non-convex. Finally, a double convex conjugation yields the convex envelope $\eMScvx(v,b)$ (solid red curve), which is constant on the interval between the minima of $\eMS(v,b)$.}
\includegraphics[scale=0.6]{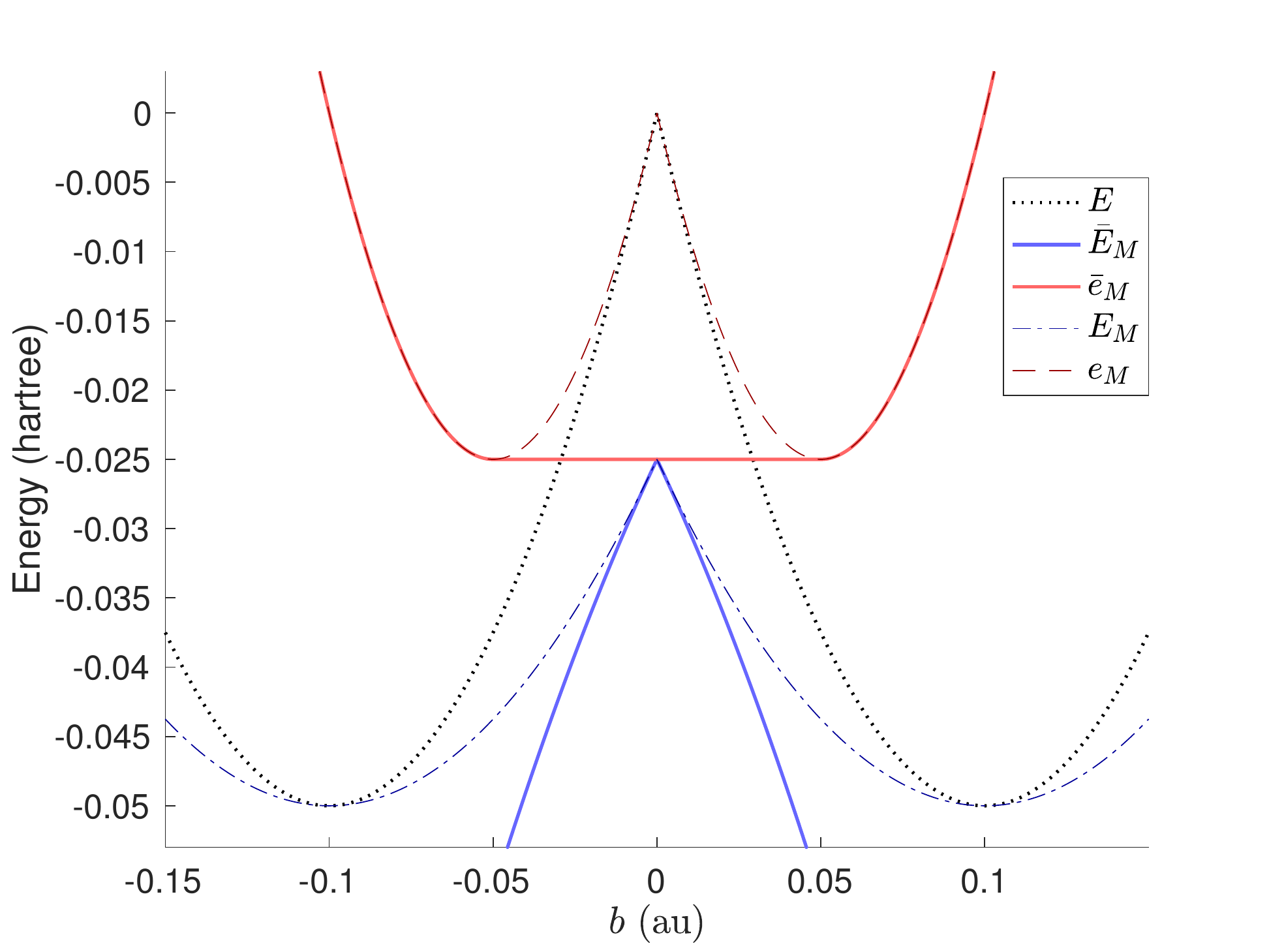}
\end{figure}

\begin{figure}[tb!]
\centering
\caption{\label{figWFSUBGRAD} On the left, the concave functional $\EMScvx(v,B)$ is displayed together with all the supergradients at two selected points (marked by circles). There is an infinite family of supergradients (dashed lines) at the non-differentiable point $B=0$. At the differentiable point $B=0.1$, the supergradient (dash-dot line) is unique. On the right, the convex functional $\eMScvx(v,\beta')$ is displayed. The minimum is attained on a finite interval (black figure). The unique subgradient at $\beta'=0.1$ is also displayed (dash-dot line). The correspondence between the infinite family of subdifferentials at the maximum of $\EMScvx$ and the unique subgradient at the minimum of $\eMScvx$ is an example of the Hohenberg--Kohn-like mapping that exists in the formalism. Another example is the correspondence between the supergradient of $\EMScvx(v,0.1)$ and the subgradient of $\eMScvx(v,0.1)$.}
\includegraphics[width=0.48\textwidth]{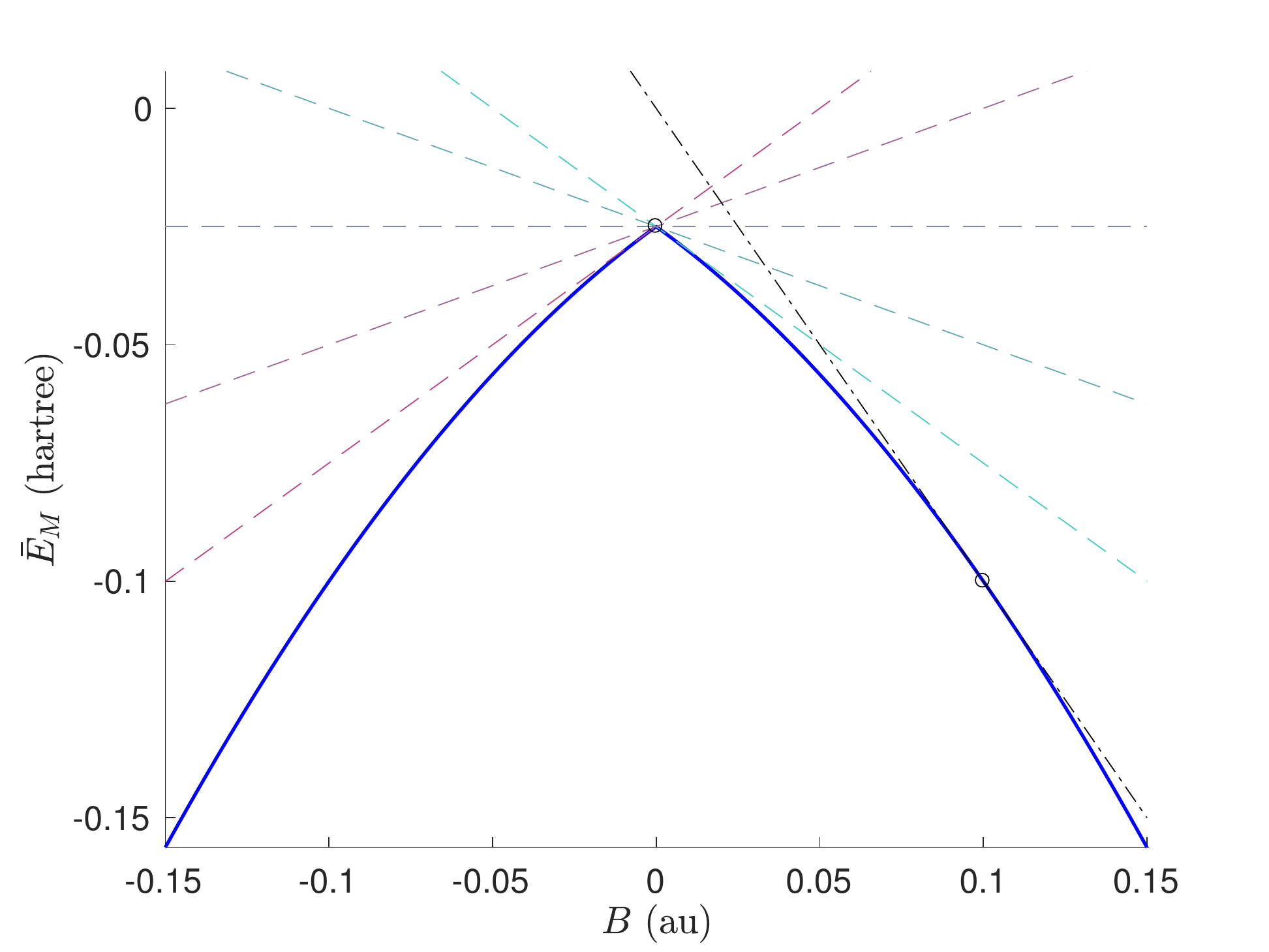}
\includegraphics[width=0.48\textwidth]{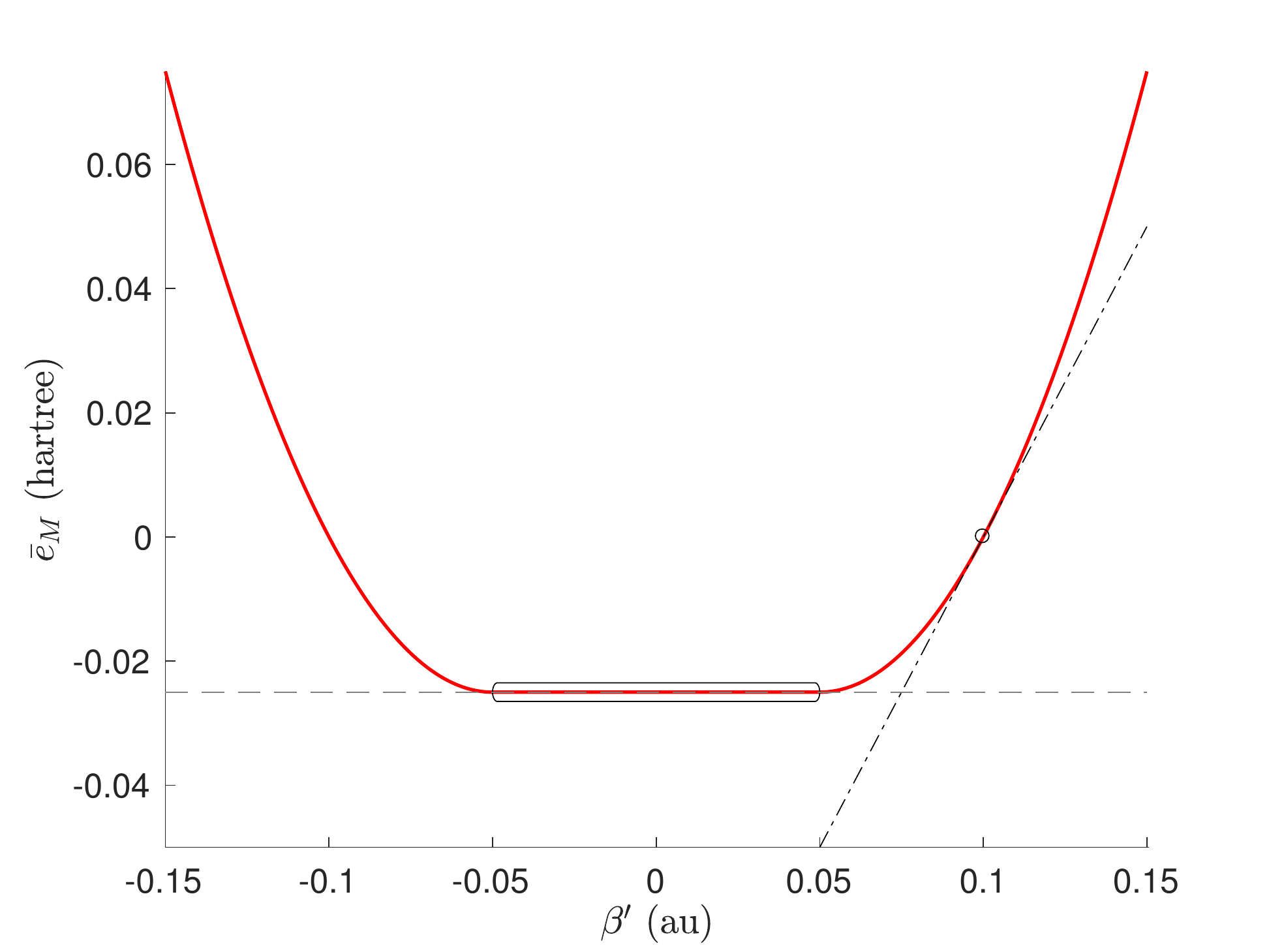}
\end{figure}

A numerical example, obtained by setting $E(v,0)=0$, $\xi = 5$, $\gamma = 1$, and $\nu = 0.1$, is shown in Fig.~\ref{figWFATOM}. Except for the cusp at $b=0$, all functionals are differentiable. The optimization problems
\begin{align}
   \EMScvx(v,B) & = \eMScvx^{\bigtriangleup}(v,B) = \inf_{\beta'} \big( -\frac{\beta' B}{\nu} + \eMScvx(v,\beta') \big),
            \\
   \eMScvx(v,\beta') & = \EMScvx^{\bigtriangledown}(v,\beta') = \sup_{B} \big( \EMScvx(v,B)  + \frac{\beta' B}{\nu} \big),
\end{align}
are therefore solved when
\begin{align}
    \frac{\partial \eMScvx(v,\beta')}{\partial \beta'} -\frac{B}{\nu} & = 0,
            \\
   \frac{\partial \EMScvx(v,B)}{\partial B}  + \frac{\beta'}{\nu} & = 0.
\end{align}
These equations establish mapping between external fields $B$ and total fields $\beta'$ that is an instance of a Hohenberg--Kohn-like mapping. The solution pair $B=0.1$ and $\beta' = 0.1$ is illustrated in Fig.~\ref{figWFSUBGRAD}. At $B=0$, the mapping becomes many-to-one. At $B=0$, the supergradient is a finite interval,
\begin{equation}
   \overline{\partial} \EMScvx(v,B=0) = \{\beta' \, | \, |\beta'| \leq \gamma/(1+2\nu\gamma)\}.
\end{equation}
This interval of total fields is mapped onto the unique external field in the subgradients
\begin{equation}
   \underline{\partial} \eMScvx(v,\beta') = \{0\}, \quad \text{for all}\ |\beta'| \leq \gamma/(1+2\nu\gamma).
\end{equation}

\subsection{Example: closed-shell paramagnetism}

\begin{figure}[tb!]
\centering
\caption{\label{fig2STATE} Functionals related to the two-state model. MY regularization of the non-convex $\Estd(v,b)$ (black dotted curve) results in the paraconcave functional $\EMS(v,b)$ (blue dash-dot curve), which is non-differentiable at $b=0$. Subtraction of the magnetic field energy yields the concave functional $\EMScvx(v,b)$. Addition of the magnetic field energy to $\Estd(v,b)$ yields the functional $\eMS(v,b)$ (red solid curve), which remains non-convex. Finally, a double convex conjugation yields the convex envelope $\eMScvx(v,b)$ (red dotted curve), which is constant on the interval between the minima of $\eMS(v,b)$.}
\includegraphics[scale=0.62 ]{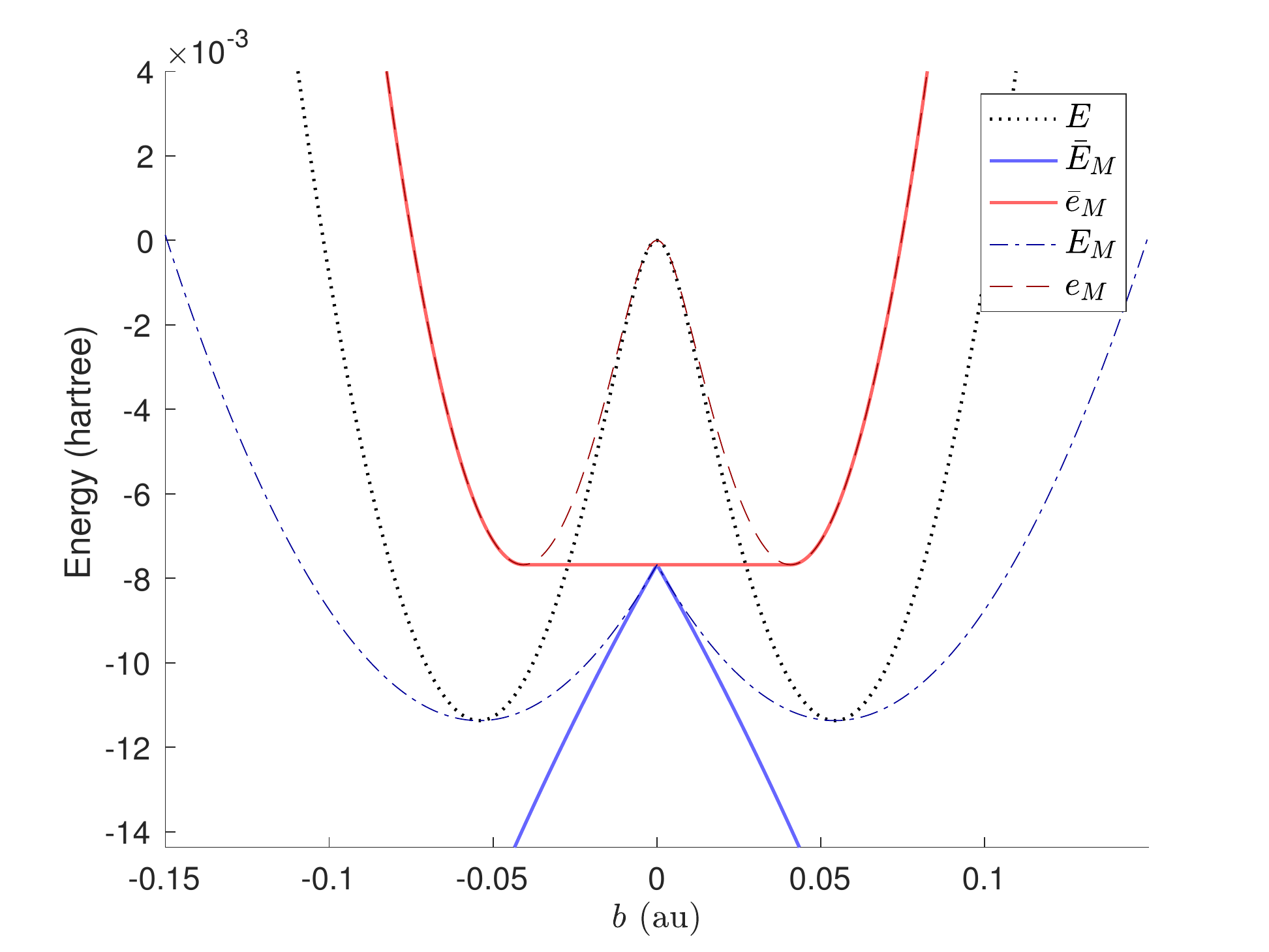}
\end{figure}

\begin{figure}[tb!]
\centering
\caption{\label{fig2STATENU} MY regularizations $M_{\nu} \Estd(v,B)$ of the two-state model energy $\Estd(v,B)$ (solid black curve). Approximations obtained with three different values $\nu = 0.005, 0.05, 0.2$ of the regularization parameter are shown (blue dash-dot curves). Smaller values yield closer approximations and $\nu = 0.005$ is sufficiently conservative to preserve the finite, negative curvature around $B = 0$.}
\includegraphics[scale=0.62]{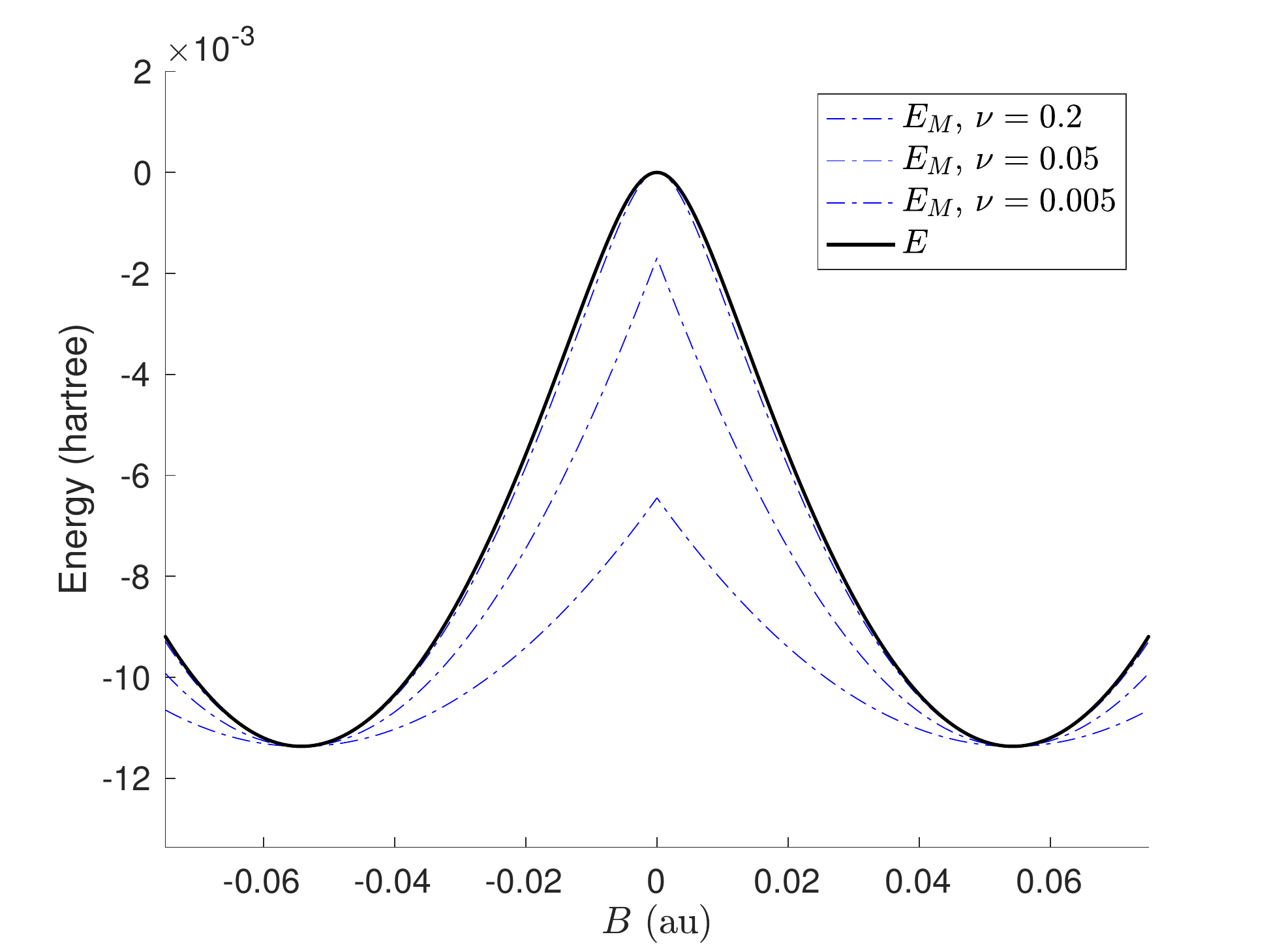}
\end{figure}

In closed-shell paramagnetic systems, the permanent magnetic moment vanishes and there is only an induced magnetic moment. Moreover, the energy initially decreases with the magnetic field strength. A paradigmatic example is the boron monohydride molecule, which is paramagnetic with respect to magnetic fields perpendicular to the bond axis. For weak and intermediate field strengths, the magnetic-field variation of the energy is well described by the following two-state model Hamiltonian~\cite{TELLGREN_PCCP11_5489},
\begin{equation}
   H = \begin{pmatrix} E_0 - \tfrac{1}{2} \chi_0 B^2 & -i m \\ i m & E_1 - \tfrac{1}{2} \chi_1 B^2 \end{pmatrix},
\end{equation}
with suitably chosen values for the model parameters $E_0$, $E_1$, $m$, $\chi_0$, and $\chi_1$. The numerical example $E_0=0$, $E_1=0.01$, $m=0.6$, $\chi_0=-14$, and $\chi_1=-8$ captures the essential qualitative features---paramagnetism for weak fields and eventually a transition to diamagnetism. The effects of varying scalar potential or density are not captured---$v$ is held fixed at the potential that corresponds to the electrostatic attraction to the boron and hydrogen nuclei. The resulting energy curve, $E(v,b)$, is shown in Fig.~\ref{fig2STATE} together with related energy functionals obtained with a large value of the regularization parameter $\nu = 0.3$ in order to make the effects visible. The fact that the Maxwell--Schr\"odinger energy $\EMS(v,b)$ has a non-differentiable cusp at $b=0$ illustrates that a MY regularization of a non-convex function sometimes results in less regularity. Moreover, the curvature of the energy $\EMS(v,b)$ near $b=0$ has the opposite sign compared to $\Estd(v,b)$. This occurs because the state with vanishing internal field $\beta = 0$ is unstable and it is energetically favorable for the system to spontaneously break time-reversal symmetry by developing an internal field. This effect only occurs for large enough $\nu$ so that
\begin{equation}
   \frac{1}{\nu} + \frac{\partial^2 \Estd(v,B)}{\partial B^2} \Big|_{B=0} < 0.
\end{equation}
Hence, the value of the regularization parameter also sets a limit on the maximum negative curvature that can occur in $\Estd$ without resulting in cusps in $\EMS$. In Fig.~\ref{fig2STATENU}, the effect on $\EMS$ of decreasing the regularization parameter is illustrated. A value of $\nu = 0.005$ is sufficiently small to avoid misrepresenting the sign of the curvature around $b=0$.

\section{Discussion and conclusion}
\label{secCONCL}

The MDFT formalism detailed above constitutes an alternative to the two most well-established extensions of standard density functional theory that accomodate external magnetic fields---BDFT~\cite{GRAYCE_PRA50_3089} and paramagnetic CDFT~\cite{VIGNALE_PRL59_2360}. It is also beneficial from the perspective that a non-relativistic DFT should be a rigorous limit of a relativistic DFT as it does not rely on the paramagnetic current density nor the change of variables $u = v + \tfrac{1}{2} A^2$. The key insight behind MDFT is that replacing the standard energy functional by the energy of a Maxwell--Schr\"odinger model leads to a paraconcave energy functional $\EMS(v,\mathbf{A})$. Moreover, subtraction of a function with the physical interpretation as the magnetic self-energy $\frac{1}{2\mu} \pair{\mathbf{B}}{\mathbf{B}}$ leads to a concave functional. Universal functionals of $(\rho,\boldsymbol{\beta}')$---or, equivalently, of $(\rho,\boldsymbol{\alpha}')$ or $(\rho,\boldsymbol{\kappa}')$---have been constructed both as constrained-search functionals and as Legendre--Fenchel transformations. The latter generalize the previous convex formulation due to Lieb~\cite{LIEB_IJQC24_243} and the cycle of partial transforms detailed in Sec.~\ref{secMDFT_LIEB} and Fig.~\ref{figMDFTrel} parallels the corresponding cycle for paramagnetic CDFT in Sec.~\ref{secPARACDFT} and Fig.~\ref{figpCDFTrel} (see also Ref.~\onlinecite{REIMANN_JCTC13_4089}). Each transformation is a variational principle that the Maxwell--Schr\"odinger energy and related MDFT functionals satisfy.

As a complement to the previous discovery that a change of variables can bring out hidden convexity properties of $E(v,\mathbf{A})$~\cite{TELLGREN_PRA86_062506}, the  results in Sec.~\ref{secMDFT_MY} show that a MY regularization brings out a different, rich set of convexity properties. The fact that the energy functionals are MY and LL regularizations of the standard Schr\"odinger energy also enables a reinterpretation of the theory and a rationale for using other values of $\mu$ than the actual vacuum permeability: $\EMS(v,\mathbf{B})$ is the closest approximation to $E(v,\mathbf{B})$ when curvature (w.r.t.\ $\mathbf{B}$) is not allowed to exceed $1/\mu$. Hence, the parameter $\mu$ can be used to choose a trade off between regularity (maximum curvature) and accuracy. When generalized to a two-parameter family of LL regularizations, as in Sec.~\ref{secMDFT_DIFF}, the universal functional is even a differentiable functional of the magnetic field. This complements the previous work by Kvaal, et al., on MY regularization with respect to the scalar potential and density variables~\cite{KVAAL_JCP140_18A518}.

A direct consequence of the convex formulation is the existence of a Hohenberg--Kohn-like mapping between (possibly degenerate) density pairs $(\rho,\boldsymbol{\kappa}')$ and potential pairs $(v,\mathbf{B}')$. Moreover, the fact that this mapping exists for all parameter values $\mu > 0$ sheds light on the status of Hohenberg--Kohn-like results for physical current densities within the Schr\"odinger model: counter-examples, if they exists at all, likely rely on pathological properties of $E(v,\mathbf{B})$ that cannot be well approximated by any MY regularization with bounded curvature. A recent independent work has established a comparable Hohenberg--Kohn-like mapping within the non-relativistic QED setting, where the electromagnetic fields are quantized~\cite{RUGGENTHALER_1509_01417}.

Finally, practical experience with available approximate paramagnetic CDFT functionals has shown that vorticity is a numerically difficult quantity to work with~\cite{ZHU_JCP125_094317,ZHU_PRA90_022504,TELLGREN_JCP140_034101}. A formulation in terms of physical currents is therefore an interesting alternative avenue to explore in the construction of practical density-functional approximations.

\section*{Acknowledgments}

This work was supported by the Norwegian Research Council through the Grant No.~240674 and CoE Centre for Theoretical and Computational Chemistry (CTCC) Grant No. 179568/V30. The author thanks S.~Kvaal, A.~Laestadius, and T.~Helgaker for useful discussions.

\newcommand{\noopsort}[1]{} \newcommand{\printfirst}[2]{#1}
  \newcommand{\singleletter}[1]{#1} \newcommand{\switchargs}[2]{#2#1}

\newpage

\end{document}